\documentclass[11pt,british]{paper}
\usepackage{geometry}
\usepackage{authblk}

\usepackage[utf8]{inputenc}
\usepackage[T1]{fontenc}
\usepackage{hyperref}
\hypersetup{colorlinks, allcolors=red, linktoc=page}
\usepackage{amsmath,revsymb,amssymb,mathtools}
\usepackage{amsthm}
\usepackage{nicematrix}

\usepackage{float}
\usepackage{physics}
\usepackage{dsfont}

\renewcommand{\emph}{\textit}

\newcommand{\iDI}[1]{{#1}}
\newcommand{\iOP}[1]{\widetilde{{#1}}}
\newcommand{\fDI}[1]{{#1}'}
\newcommand{\fOP}[1]{\widetilde{{#1}'}}

\usepackage[
    sorting=none,
    maxbibnames=99,
    url=false
]{biblatex}
\usepackage[most]{tcolorbox} 
\definecolor{myblue}{rgb}{0.6,0.6,0.9}
\newtcolorbox{examplebox}[1]{
    enhanced,          
    breakable,  
    colback=myblue!5!white,    
    colframe=myblue,           
    width=\textwidth,          
    arc=2pt,                   
    boxrule=1.5pt,             
    top=12pt,
    bottom=10pt,
    left=12pt,
    right=12pt,
    title={#1},                
    coltitle=black,           
    fonttitle=\bfseries,       
    attach boxed title to top left={yshift=-2mm, xshift=2mm},
    boxed title style={
        sharp corners, 
        size=small, 
        colback=myblue!40,    
        colframe=myblue!40,  
    },
    before skip=18pt,
    after skip=12pt,
}
\definecolor{mygreen}{RGB}{255, 51, 51}
\newtcolorbox{recipebox}[1]{
    enhanced,          
    breakable,  
    colback=mygreen!5!white,    
    colframe=mygreen,           
    width=\textwidth,          
    arc=2pt,                   
    boxrule=1.5pt,             
    top=12pt,
    bottom=10pt,
    left=12pt,
    right=12pt,
    title={#1},                
    coltitle=black,           
    fonttitle=\bfseries,       
    attach boxed title to top left={yshift=-2mm, xshift=2mm},
    boxed title style={
        sharp corners, 
        size=small, 
        colback=mygreen!40,    
        colframe=mygreen!40,  
    },
    before skip=18pt,
    after skip=12pt,
}

\makeatletter
\renewcommand{\l@section}[2]{%
  \vspace{0.5em}
  \@dottedtocline{1}{1.5em}{2.3em}{#1}{#2}}
\makeatother
\makeatletter
\renewcommand{\l@subsection}[2]{%
  {\@dottedtocline{2}{2.8em}{2em}{#1}{#2}}}
\makeatother

\makeatletter
\renewcommand{\l@subsubsection}[2]{} 
\makeatother



\begin{document}

\title{\Large{Systematic derivation of Tsirelson bounds in arbitrary dimensions}}

\author[1]{\large{Lorenzo Coccia}\thanks{lorenzo.coccia@unipd.it} \ }
\author[1]{Matteo Padovan}
\author[1,2]{Giuseppe Vallone}

\affil[1]{\small{Dipartimento di Ingegneria dell'Informazione, Universit\`a degli Studi di Padova, via Gradenigo 6B, IT-35131 Padova, Italy}}
\affil[2]{\small{Padua Quantum Technologies Research Center, Universit\`a degli Studi di Padova, via Gradenigo 6B, IT-35131 Padova, Italy}}

\maketitle

\begin{abstract}
   The study of Bell nonlocality and the bounds of quantum correlations, the so-called {\it Tsirelson bounds}, is fundamental to quantum information science and the exploration of the limits of quantum theory. 
   While quantum bounds for qubit systems have been extensively characterized, determining tight quantum bounds for correlations attainable with high-dimensional quantum states and measurements remains a significant challenge. 
   In this work, we propose a systematic derivation of bipartite 
   Tsirelson and local bounds written in terms of sum-of-squares decompositions. Using this method, we discover novel bounds and recover established results for maximally entangled states of qubits and qu$d$its.
\end{abstract}

\tableofcontents

\section{Introduction}
In 1964, Bell's celebrated paper \cite{Bell:1964kc} transformed the long-standing debate on the completeness of quantum mechanics into a quantitative and experimentally testable problem. 
By deriving inequalities that must be satisfied by any theory based on local hidden variables (LHV), Bell showed that the predictions of quantum mechanics could, in principle, be distinguished from those of any locally causal model through statistical correlations observed in experiments \cite{Brunner2014}.

Although originally introduced to rule out local hidden-variable descriptions of nature, Bell inequalities have gradually evolved into an essential practical resource. 
They now constitute the key ingredient of device-independent quantum information processing, where one aims to infer properties of quantum systems from observed measurement statistics \cite{scarani2012device, Supic2020, Zapatero2023, Primaatmaja2023securityofdevice}. 
In this paradigm, tasks such as the generation of certified randomness or the establishment of secure cryptographic keys can be guaranteed without relying on the internal workings of the quantum devices employed.

In a Bell scenario, several spatially separated parties perform measurements on subsystems of a shared physical resource and record the outcomes. 
Each party chooses from a set of possible measurements (inputs), and the experiment produces a joint probability distribution of outcomes conditioned on these choices. 
The nature of the physical theory underlying the experiment constrains the set of achievable correlations. 
For instance, correlations admitting a local hidden variables description must lie within a polytope defined by convex combinations of deterministic strategies \cite{Fine1982, Brunner2014}. 

The hyperplanes bounding the local polytope define Bell inequalities, namely linear constraints that all LHV correlations must satisfy. 
While quantum mechanics allows for correlations that extend beyond this polytope and violate these inequalities, quantum correlations are themselves confined to a broader, non-polytopic convex set.
The bound of a Bell inequality violation permitted by quantum theory is known as the quantum or Tsirelson bound, named after the seminal work of Boris S. Tsirelson \cite{Cirelson:1980ry}.
Determining these bounds not only quantifies the gap between classical and quantum correlations but also reveals fundamental structural properties of quantum theory. 
A remarkable example is self-testing \cite{Supic2020}: certain correlations residing on the border of the quantum set can essentially be performed by only one possible set of measurements and states (up to change of basis and auxiliary degrees of freedom). 
This is particularly relevant for certifying the correct operation of a device without the need to characterize it. 
Consequently, self-testing results are leveraged in many device-independent quantum security protocols across various scenarios (see e.g. \cite{Padovan2026, Coccia2026, Padovan2024, Foletto2021, Farkas2026, Acin2016}).

Understanding the joint geometry of local and quantum bounds is, however, non-trivial, and our current knowledge remains limited \cite{OpenQuantumProblems_IQOQI}. 
A "minimal" Bell scenario involving two users, each performing two measurements with two outputs (the $(2,2,2)$ scenario), is relatively well understood: all extremal points of the quantum correlations set have been characterized 
\cite{
Tsirelson:1987lho,Masanes2003, Le:2021dvl, Barizien2025, Barizien:2024uab}, and the literature offers a wide variety of Tsirelson bounds and techniques \cite{Barizien_2024, Coccia2026} to obtain them using both maximally and non-maximally entangled states.

In higher-dimensional systems, as the number of measurement outcomes increases, the complexity of the quantum set scales exponentially, and analytical methods for systematically deriving sum-of-squares decompositions remain scarce. 
For instance, while the approach in \cite{Barizien_2024} successfully derives sum-of-squares decompositions tailored to specific shared quantum states, it becomes increasingly difficult to apply as the Hilbert space dimension grows. 
In these complex scenarios, numerical approaches,  like those based on Navascués-Pironio-Acín (NPA) hierarchy \cite{navascues_bounding_2007,navascues_convergent_2008, Mortimer:2025gxw}, are possible. 
However, numerical methods offer limited insight into the underlying analytical structure or the physical principles defining the quantum set. 
Furthermore, the NPA approach can become computationally intractable at higher levels \cite{Gigena2025} or finite levels of the hierarchy may fail to converge tightly to optimal quantum bounds in specific scenarios \cite{fanizza2025}.
For these reasons, this work investigates an analytical approach to the study of the set of quantum correlations, contributing to the systematic characterization of the quantum boundary for correlations achieved using bipartite maximally entangled states in arbitrary dimensions. 

From a mathematical standpoint, analytically proving Tsirelson bounds often relies on the technique of sum-of-squares (SOS) decompositions \cite{Supic2020}. 
The idea is to rewrite the Bell operator $S$ associated with a given inequality as a sum of positive semidefinite terms: $S = \sum_i P_i^\dagger P_i$. The positivity of each term $P_i^\dagger P_i$ immediately implies that the expectation value of $S$ on any quantum state is lower-bounded by $0$, thus establishing a bound. Moreover, the quantum states that achieve this bound must satisfy the constraints $P_i \ket{\psi}=0$ for all $i$. This can be exploited to characterize the structure of the optimal state $\ket{\psi}$ and the corresponding measurements in self-testing \cite{Supic2020, Barizien_2024, Barizien2025, Gigena2025, Sarkar2021}. In this way, the SOS method provides a powerful bridge between observed correlations and the analytical conditions required to derive robust self-testing results.

Our method builds upon these ideas and is detailed in Section \ref{sec:gen_disc}. 
We begin with a "trivial" Bell operator, defined via a sum-of-squares decomposition, for which the local and quantum bounds coincide. 
For this baseline operator, an initial optimal strategy can be readily identified, involving a maximally entangled state and arbitrary measurements by one of the parties.
The core of our approach is to then apply specific transformations to this initial operator to derive a new class of physically observable Bell operators.
Crucially, we find that these transformations are a subset of unitary transformations, ensuring that the resulting Bell operator is physically observable, and constrained such that an optimal quantum strategy remains identifiable.
In particular, the new optimal strategy is derived directly from the initial strategy and the applied transformation.
The family of SOS Bell operators we derive have the following simple structure:
\begin{equation}
\label{eq:SOS_intro}
    S = \mathbf{G}^\dag \mathbf{G}\,,\qquad  \mathbf{G}=M{\bf \Lambda}-U\Gamma{\bf \Pi} \,.
\end{equation}
This expression can be interpreted as follows. 
The vectors ${\bf \Lambda}$ and ${\bf \Pi}$ contain the symbolic local projectors for Alice and Bob, respectively. 
The matrices $M$ and $\Gamma$ characterize how the projectors for a given measurement enter the SOS Bell operator for Alice and Bob. 
Meanwhile, the unitary matrix $U$ mixes Bob's projectors across different measurements.
Because $S$ is a sum-of-squares decomposition, it defines a boundary $\langle S \rangle \geq 0$. 
This bound is tight because it can be saturated by a maximally entangled state using a known quantum measurement strategy;
thus, the condition $\langle S \rangle \geq 0$ defines a valid Tsirelson bound. 
Conversely, the local bound is generally strictly greater than zero, and can be computed by standard numerical optimization or checking over the finite set of deterministic strategies, as we will discuss.

Within the compact formalism of Eq.~\eqref{eq:SOS_intro}, our approach yields Bell operators that are saturated by a maximally entangled state and a \emph{generic} choice of measurements (projectors) on Alice’s side, for an \emph{arbitrary} number of inputs and outcomes. 
As special cases, we recover a large class of Bell inequalities for qubits, including those introduced in \cite{Christensen2015, Wooltorton2022, Barizien_2024}, as well as the elegant Bell inequality of \cite{Gisin:2007gps}.
For qudits, we first reproduce the Bell inequality for mutually unbiased bases (MUBs) introduced in \cite{Tavakoli:2020zlz} and derive a new inequality for MUBs requiring fewer inputs. 
We then reproduce and generalize the SATWAP inequality with two inputs per party \cite{SATWAP2017}.

The paper is organised as follows. 
In Sec.~\ref{sec:gen_disc}, we introduce the mathematical formalism and the analytical method proposed in this work. At the end of this section, we summarize the procedure in a highlighted red box.
Sec.~\ref{sec:qubit_strate} is devoted to applying our method to derive quantum bounds saturated by qubit strategies. 
Previously known results from the literature are reproduced in highlighted blue boxes.
In Sec.~\ref{sec:high_dim}, we discuss Bell operators associated with high-dimensional, $d >2$ strategies.
We first discuss scenarios in which Alice has two inputs and Bob $2d$ inputs, with specific examples for the MUBs, and then we consider a symmetric scenario in which both parties have two inputs. 
Finally, in Sec.~\ref{sec:conclusions}, we present the outlook and discuss potential future directions of this work.

\section{Description of the method}
\label{sec:gen_disc}

Consider a standard Bell scenario involving two spatially separated parties, Alice and Bob. Each party chooses from $m$ possible measurement settings, denoted by $x, y \in \{1, \dots, m\}$, which yield one of $d$ possible outcomes, $a, b \in \{1, \dots, d\}$. The experimental statistics are characterized by the joint probability distribution $p(a, b | x, y)$, referred to as correlations.

In the quantum mechanical framework, we model this scenario by assigning Alice and Bob $m$ sets of $d$ orthogonal projectors $
\Lambda_a^{(x)}$ and $\Pi_b^{(y)}$ acting on the Hilbert spaces $\mathcal{H}_A$ and $\mathcal{H}_B$,
respectively. These projectors satisfy the following orthogonality and completeness relations:
\begin{equation}
\label{eq:normalization_cons}
\iDI{\Lambda}_a^{(x)} \iDI{\Lambda}_{a'}^{(x)} = \delta_{a a'} \iDI{\Lambda}_a^{(x)}, \quad \sum_{a=1}^{d} \iDI{\Lambda}_a^{(x)} = \openone, \quad \forall x
\end{equation}
and similarly for $\Pi^{(y)}_b$. 
The observed correlations are obtained as the expectation value of these local operators acting on a shared quantum state $\ket{\psi} \in \mathcal{H}_A \otimes \mathcal{H}_B$:
\begin{equation}
\label{eq:correlations}
p(a,b|x,y) = \expval{\iDI{\Lambda}^{(x)}_a \otimes \iDI{\Pi}^{(y)}_b}{\psi} \, .
\end{equation}

Bell operators are observables whose expectation values are linear combinations of the correlations $p(a, b | x, y)$. 
Since the range of their attainable expectation values may differ between local realistic theories and quantum theory, they can serve as witnesses that certify which physical frameworks are compatible with experimental data \cite{Brunner2014}. 

Identifying the theoretical boundaries of the expectation values of a Bell operator $S_0$ is greatly simplified if $S_0$ is written as sum-of-squares (SOS) decomposition:
\begin{equation}
\label{eq:SOSform}
S_0 = \sum_{i=1}^N P_i^\dagger P_i\,,
\end{equation}
where $N$ is an integer and each $P_i$ is a function of Alice’s and Bob’s local projectors, $P_i(\Lambda_a^{(x)}, \Pi^{(y)}_b)$. 
The primary advantage of this decomposition is that the resulting operator is explicitly positive semi-definite. 
Consequently, if a specific quantum strategy yields $\langle S_0 \rangle = 0$, then zero represents a Tsirelson bound. 
This quantum bound can then be compared to the classical limit, which is necessarily greater than or equal to zero.

In this work, we aim to develop a systematic method to construct Bell operators in the SOS form \eqref{eq:SOSform}.
These operators must satisfy two requirements:
\begin{itemize}
    \item \textit{Observability}:  By definition, the expectation value of a Bell operator must be expressible as a linear combination of the observed correlations \eqref{eq:correlations}. This implies that $S_0$ cannot contain products of operators from the same party with different inputs, such as $\Lambda^{(x)}_a \Lambda^{(x')}_{a'}$ for $x \neq x'$. 
    Such "cross-terms" represent quantities that are not directly observable in a standard Bell scenario.
    \item \textit{Saturation}: There must exist a quantum optimal strategy (a state $\ket*{\widetilde{\psi}}$ and measurements $\iOP{\Lambda}^{(x)}_a$ and $\iOP{\Pi}^{(y)}_b$) that saturates the bound $\expval{S_0}=0$. This implies that $P_i(\widetilde{\Lambda}_a^{(x)},\widetilde{\Pi}^{(y)}_b)\ket*{\widetilde{\psi}}=0$  for all $i$.
\end{itemize}
In the remainder of this paper, we derive Bell operators whose Tsirelson bounds are saturated using maximally entangled states $\ket*{\widetilde{\psi}}=\ket*{\Phi^+}$. 
Our approach involves two steps: first, we identify a simple initial Bell operator and its optimal strategy; 
Subsequently, we apply transformations to construct families of general operators.

A comment on notation: Throughout this paper, we denote the specific operators of a quantum strategy with a tilde ($\tilde{\Lambda}$), whereas the abstract symbolic operators used to define the Bell operator are written without a tilde ($\Lambda$).

\subsection{A simple initial operator}
A particularly simple way to fulfill the previous requirements is the following.
\begin{itemize}
    \item To satisfy the \textit{observability} constraint, we define each $P_i$ as a function of only a single input $x$ for Alice and $y$ for Bob:
\begin{equation}
\label{eq:no_mix_input}
    P_i \equiv P^{(xy)}_i = \sum_{a=1}^d p_{i,a}^{(x)}\Lambda_a^{(x)} + \sum_{b=1}^d q_{i,b}^{(y)}\Pi^{(y)}_b
\end{equation}
for some coefficients ${p_{i,a}^{(x)}, q_{i,b}^{(y)}}$. 
This structure ensures that the resulting SOS decomposition contains no cross-terms between different measurement settings of the same party.
\item To satisfy the \emph{saturation} condition, we exploit a property of maximally entangled states,
\begin{equation}
     \ket*{\Phi^+}=\frac{1}{\sqrt{d}}\sum_i \ket{i}\ket{i} \ . 
\end{equation}
For any choice of projectors $\widetilde{\Lambda}_a^{(x)}$ on Alice's side, Bob's projector $\widetilde{\Pi}^{(x)}_a=(\widetilde{\Lambda}^{(x)}_a)^T$, where the transpose is computed in the computational basis $\ket{i}$, satisfies (see e.g.\ Lemma 4.4 in \cite{Paddock:2023bnh})
\begin{equation}
\label{eq:cond_max_state}
\qty(\widetilde{\Lambda}_a^{(x)} \otimes \openone-\openone \otimes \widetilde{\Pi}_a^{(x)}) \ket*{\Phi^+}=0.
\end{equation} 

The above points tells us that if we consider the pieces of the SOS decomposition to be
\begin{equation}
P_{i_x}^{(x)} = \sum_{a=1}^{d} M^{(x)}_{i_x, a} \qty(\Lambda_a^{(x)} - \Pi_a^{(x)})\,,\qquad
i_x=1,\cdots, n_x
\end{equation}
where $M^{(x)}$ are generic $n_x \times d$ complex matrices, the bound $\langle S_0 \rangle = 0$ is reached by choosing a maximally entangled state, $\ket*{\widetilde{\psi}}=\ket*{\Phi^+}$, arbitrary sets of projectors  $\widetilde{\Lambda}_a^{(x)}$ on Alice's side and $\widetilde{\Pi}_a^{(x)} = (\widetilde{\Lambda}^{(x)}_a)^T$ on Bob's.
Each $P^{(x)}_{j_x}$ represents a
combination of projectors belonging
to the same input $x$, while $n_x$ is the
number of such combinations for any input $x$. 
\end{itemize}
Hence a simple Bell operator involving all the projectors and satisfying our requirements is
\begin{equation}
\label{eq:basic_inequality}
    S_0 =\sum_{x=1}^{m}\sum_{i_x=1}^{n_x}\qty(P^{(x)}_{i_x})^\dag P^{(x)}_{i_x}\,.
\end{equation}
To match our previous notation in Eq.~\eqref{eq:SOSform}, we define $N=\sum_x n_x$ and introduce a global index $i \in \{1,\dots,N\}$. 
This single index replaces the nested structure, running over all different inputs and absorbing the individual sub-indices $i_x=1,\dots,n_x$ associated with each input $x$.

The condition $\expval{S_0} = 0$ defines a Tsirelson bound, i.e., a limit on the correlations attainable within quantum mechanics. 
However, this bound is not particularly interesting, as it can also be saturated by local realistic strategies. 
In this sense, $\expval{S_0} \ge 0$ constitutes both a Tsirelson inequality and a Bell inequality, making the Bell operator $S_0$ unsuitable for probing nonlocal features of quantum mechanics or for technological applications exploiting nonlocality.
Nevertheless, one may attempt to transform \eqref{eq:basic_inequality} in order to derive less trivial Bell operators. 
To this end, we collect the projectors into $md$-dimensional vectors
\begin{equation}
\begin{aligned}
  {\bf \iDI{\Lambda}} &=
  \qty(\iDI{\Lambda}^{(1)}_1 \cdots \iDI{\Lambda}^{(1)}_{d} \cdots
   \iDI{\Lambda}^{(m)}_1 \cdots \iDI{\Lambda}^{(m)}_{d})^T,
  \\
  {\bf \iDI{\Pi}} &=
  \qty(\iDI{\Pi}^{(1)}_1 \cdots \iDI{\Pi}^{(1)}_{d} \cdots
   \iDI{\Pi}^{(m)}_1 \cdots \iDI{\Pi}^{(m)}_{d})^T ,
\end{aligned}
\end{equation}
where the transpose is taken on the $md$-dimensional space and not on the Hilbert space.
With this notation, the sum-of-squares operator \eqref{eq:basic_inequality} can be written compactly as
\begin{equation}
\label{eq:SOS_M}
    S_0=({\bf \iDI{\Lambda}-\iDI{\Pi}})^\dagger M^{\dagger}M({\bf \iDI{\Lambda}-\iDI{\Pi}})\ , 
\end{equation}
 where we introduced a block-diagonal matrix $ M=\bigoplus_x M^{(x)}$. 
Note that the block diagonal structure allows to maintain the observability.

The central idea of this work is to apply suitable transformations to obtain less trivial Bell operators.
Such operators will be again written in terms of an SOS decomposition, and for this reason the Tsirelson bound will be always fixed to 0, while the local bound will in general be larger than 0.
Since our technique is constructed around the relation \eqref{eq:cond_max_state}, the new optimal strategies will involve the maximally entangled state.
While analogous relations exist for more general entangled states, their application to constructing Bell operators is more challenging (see Sec.\ \ref{sec:conclusions}).

\subsubsection{SOS decompositions in terms of normal operators}
\label{sec:normal_op}
Before describing our results, it is useful to  discuss in more detail the expression \eqref{eq:SOS_M}, in particular the role and meaning of the matrix $ M$.
For this purpose we introduce the operators
\begin{align}
\mathbf{\iDI{A}}
&
\equiv M \mathbf{\iDI{\Lambda}}\,,\qquad \iDI{A}_{i_x}^{(x)} = \sum_{a}M^{(x)}_{i_x a}\iDI{\Lambda}_{a}^{(x)} \ , 
\\
\mathbf{\iDI{B}}
&
\equiv M \mathbf{\iDI{\Pi}}\,,\qquad \iDI{B}_{i_y}^{(y)} = \sum_{b}M^{(y)}_{i_y b}\iDI{\Pi}_{b}^{(y)} \ .
\end{align} 
In these definitions we are combining together the projectors of a fixed input, $x$ for Alice and $y$ for Bob, to define sets of normal operators which satisfy the following commutation relations:
\begin{equation}
    \left[\iDI{A}_{i_x}^{(x)}, {\iDI{A}_{i_x'}^{(x)}}^\dagger\right] = 
    \left[\iDI{B}_{i_x}^{(x)}, 
    {\iDI{B}_{i_x'}^{(x)}}^\dagger\right] = 0 \,, 
    \qquad \forall i_x, i'_x
\end{equation}
as it can be easily checked using the orthogonality relations \eqref{eq:normalization_cons} of the projectors.
Being normal, for each $x$ and $y$ these operators can be simultaneously diagonalized for any given input and expressed as linear combinations of the original projectors. 
Note also that, for each $x$ and $i_x$, the operators $A^{(x)}_{i_x}$ and $B^{(x)}_{i_x}$ share the same eigenvalues.
These definitions turns \eqref{eq:SOS_M} into the diagonal form
\begin{equation}
\begin{split}
    S_0&= \sum_{x=1}^m\sum_{j_x=1}^{n_x} \left( \iDI{A}_{j_x}^{(x)} - \iDI{B}_{j_x}^{(x)} \right)^\dagger \left( \iDI{A}_{j_x}^{(x)} - \iDI{B}_{j_x}^{(x)} \right)\\
    &= 
    (\mathbf{\iDI{A}}-\mathbf{\iDI{B}})^\dagger (\mathbf{\iDI{A}}-\mathbf{\iDI{B}}) =( \mathbf{\iDI{A}}^\dagger \mathbf{\iDI{A}} - \mathbf{\iDI{A}}^\dagger \mathbf{\iDI{B}} - \mathbf{\iDI{B}}^\dagger \mathbf{\iDI{A}} + \mathbf{\iDI{B}}^\dagger \mathbf{\iDI{B}} )\ . \label{eq:SOS_AB}
\end{split}
\end{equation}
Using the index notation introduced above, we assign each vector $\mathbf{A}$ (and $\mathbf{B}$) a global index $j=1, \dots , N$ obtained by concatenating the $m$ sequences of indices $j_x$.

Parameterizing Bell operators and correlations in terms of normal operators is a well-established technique in the literature for both qubit and high-dimensional Bell inequalities \cite{Brunner2014, Supic2020, Barizien_2024, Barizien2025, Bamps2015,Wooltorton2022,Sarkar2021,SATWAP2017}. 
In our discussion, representing the Bell operators in terms of the set of normal operators $\mathbf{A}$ and $\mathbf{B}$ is a convenient way to encode the structure of the  Bell operator $S_0$ in \eqref{eq:basic_inequality}. 
In this sense, when we express the optimal strategy in terms of these operators
\begin{equation}
    \mathbf{\iOP{A}}=M{\bf\iOP{\Lambda}} \ , \qquad \mathbf{\iOP{B}}=M{\bf\iOP{\Pi}} \ , 
\end{equation}
we include both the choice of initial optimal strategy and the choice of the initial operator $S_0$.

\subsection{Obtaining non-trivial Tsirelson bounds}
\label{sec:obs_bell_ops}
The Bell operator $S_0$ written in \eqref{eq:SOS_M} defines, via its expectation value, a trivial Tsirelson bound for a scenario with $m$ inputs and $d$ outcomes per party, and it is obtained with the strong requirement \eqref{eq:no_mix_input} of having SOS decompositions of just one input in each term.
In order to obtain more general and non-trival Bell operators from Eq.\ \eqref{eq:SOS_M}, we write the initial $\mathbf{\iDI{A}}$ and $\mathbf{\iDI{B}}$ sets of (symbolic) operators by means of invertible linear transformations $V_A$ and $V_B$ and other sets of (symbolic) operators $\mathbf{\fDI{A}}$ and $\mathbf{\fDI{B}}$:
\begin{equation}
\label{eq:AVA}
    {\bf \iDI{A}} = V_A \, {\bf \fDI{A}}, 
    \qquad
    {\bf \iDI{B}} = V_B \, {\bf \fDI{B}}\,.
\end{equation}
The role of the transformations $V_{A/B}$, which we assume to be invertible, is to mix operators of different "initial" inputs in a controlled way. 
To clarify this, let's make more explicit, in the index notations, how the primed operators \eqref{eq:AVA} are defined
\begin{equation}
\begin{aligned}
    \fDI{A}_{k}&=\sum_{x=1}^m\sum_{i_x=1}^{n_x}(V^{-1}_{A})_{k,i_x}{\iDI{A}}_{i_x}^{(x)} \\
    &= \sum_{i=1}^{N}(V^{-1}_{A})_{k,i}{\bf \iDI{A}}_i
\end{aligned}
\begin{aligned}\qquad \qquad  
    \fDI{B}_{w}&=\sum_{y=1}^m \sum_{i_y=1}^{n_y}(V^{-1}_{B})_{w,i_y}{\iDI{B}}_{i_y}^{(y)} \\
    &= \sum_{i=1}^{N}(V^{-1}_{B})_{w,i}{\bf \iDI{B}}_i \ .
\end{aligned}
\end{equation}
The inner sum over $i_x$ mixes operators inside the $x$-th input while the external sum over $x$ puts together contributions from different inputs.
The resulting new operators $\fDI{A}_k$ and $\fDI{B}_w$ identify, for each $k,w=1,\dots, N$, a new input with $d$ outputs.
So, each of them must  admit a spectral decomposition in terms of its own set of projectors:
\begin{equation}
\label{eq:diagon:Aprime}
    \fDI{A}_{k}=\sum_{a=1}^d\alpha^{(k)}_{a}{\fDI{\Lambda}}^{(k)}_{a} \ , \qquad \fDI{B}_{w}=\sum_{b=1}^d\beta^{(w)}_{b}{\fDI{\Pi}}_{b}^{(w)}\,.
\end{equation}
The transformed Bell operator, defined in a new scenario with $N$ measurements per party, will be
\begin{equation}
\label{eq:trans_sos}
\begin{split}
    S
    &= {\bf \fDI{A}}^\dagger V_A^\dagger V_A {\bf \fDI{A}}
    - {\bf \fDI{A}}^\dagger V_A^\dagger V_B {\bf \fDI{B}}
    - {\bf \fDI{B}}^\dagger V_B^\dagger V_A {\bf \fDI{A}}
    + {\bf \fDI{B}}^\dagger V_B^\dagger V_B{\bf \fDI{B}} \, 
\end{split}
\end{equation}
which can be written in terms of projectors by using the relations \eqref{eq:diagon:Aprime}.
However, we still need to satisfy our constraints of observability and saturation.
\begin{itemize}
    \item \emph{Observability:}
    To ensure the observability of the Bell operator $S$, we must exclude products of operators from the same party that involve different inputs, such as $\fDI{A}_{k}\fDI{A}_{k'}$ for $k \neq k'$. 
Looking at Eq.\ \eqref{eq:trans_sos} we see that such non-observable "cross-input" terms can only originate from the terms $\mathbf{A'}^\dagger 
V_A^\dagger V_A \mathbf{A'}$ and $\mathbf{B'}^\dagger 
V_B^\dagger V_B \mathbf{B'}$.
When each $k$ corresponds to a different input, in order to preserve observability, we require $V^\dagger V$ to be diagonal for both Alice and Bob.
More general expressions for $V$ are allowed if different $k$ can be associated with the same inputs. These aspects will be discussed in detail in the next section. 
Ultimately, \textit{it is sufficient to consider $V$ to be unitary}, since this choice does not limit the families of obtainable Tsirelson bounds.

\item \emph{Saturation:} To saturate the transformed bound $\expval{S} = 0$, one might formally define an optimal strategy by considering the inverse transformations of $V_{A/B}$ applied to the initial optimal strategies $\mathbf{\iOP{A}}$, $\mathbf{\iOP{B}}$:
\begin{equation}
\label{eq:opti_strate_U}
\mathbf{\fOP{A}} = V^{-1}_A \mathbf{\iOP{A}} \ , \qquad \mathbf{\fOP{B}} = V^{-1}_B \mathbf{\iOP{B}} \,.
\end{equation}
However, it is not guaranteed a priori that these transformed operators correspond to a valid quantum strategy. 
To be a valid quantum strategy, $\mathbf{\fOP{A}}$ and $\mathbf{\fOP{B}}$ must be composed of normal operators,
\begin{equation}
\label{eq:commutator_condition}
    \qty[\fOP{A}_k, \qty(\fOP{A}_k)^\dagger] = 0, \qquad \qty[\fOP{B}_k, \qty(\fOP{B}_k)^\dagger] = 0 \quad \forall k=1, \dots,N
\end{equation}
i.e.\ each of their element must be diagonalizable.
Therefore, to ensure the transformed Bell operator possesses a well-defined quantum bound of zero, we must identify transformations $V_A$ and $V_B$ that satisfy the above commutation relations.
We note that, for real choices of the matrix $M$ (i.e., when the operators $\iOP{A}_k$ and $\iOP{B}_k$ are Hermitian), a simple solution of \eqref{eq:commutator_condition} always exists, by choosing real matrices $V_A$ and $V_B$. 
\end{itemize}

\subsection{Measurement scenarios and allowed transformations}
\label{sec:scenarios}
From the previous section, the Bell operator \eqref{eq:trans_sos} is generally associated with $N$ distinct inputs per party, one for each operator $\fDI{A}_k$ and $\fDI{B}_k$. 
However, the number of inputs can be reduced by exploiting the commutation properties of the optimal strategy that saturates the Tsirelson bound. 
For example, if $\qty[\fOP{A}_{k_1},\fOP{A}_{k_2}]=0$, there is no need to treat $k_1$ and $k_2$ as separate inputs; the Tsirelson bound can be saturated by performing a single measurement defined in terms of their common projectors. 
Consequently, even at the level of the symbolic Bell operator, we can identify $k_1$ and $k_2$ as a single input.
In this way, we can define a new Bell operator with fewer inputs, which effectively shares the same optimal strategy as the original operator. Both operators are equally valid, but their properties away from the saturation point (e.g., robustness to experimental noise or to finite-size effects) can generically be different, since the number of measurements is different.
The optimal strategy 
$\{\fOP{A}_k\}$ ($\{\fOP{B}_k\}$) tells us
the minimal scenario to consider, namely the
number $m_A$ ($m_B$) of different inputs at
Alice (Bob) side.

To formalize this concept, we first need to identify sets of mutually commuting operators within the final optimal strategy (the one saturating $\ev{S}=0$).
Suppose Alice and Bob have $m_A$ and $m_B$ such sets, labeled by indices $x$ and $y$, containing $n^A_x$ and 
$n^B_y$ elements respectively. 
Since the operators inside each set $x$ and $y$ share a common set of projectors, they can be associated with the same measurement. 
Hence, we can adopt the notation from Sec.\ \ref{sec:normal_op}, and denote an element of the sets of commuting operators as $\fDI{A}^{(x)}_{i_{x}}$ and $\fDI{B}^{(y)}_{i_{y}}$, where $i_{x}=1,\dots,
n^A_x$, $i_{y}=1,\dots,
n^B_y$, $x= 1,\dots,m_A$ and $y=1,\dots,m_B$.
By applying permutations, we can always reorder $\fDI{{\bf A}}$ and $\fDI{\bf{B}}$ to group together the operators of each set $x$ and $y$:
\begin{equation}
\begin{split}
    \fDI{{\bf A}}&=\qty(\fDI{A}^{(1)}_{1},\cdots,\fDI{A}^{(1)}_{n^A_1},\cdots,\fDI{A}^{(m_A)}_{1},\cdots,\fDI{A}^{(m_A)}_{n^A_{m_A}}) \ ,  \\
    \fDI{{\bf B}}&=\qty(\fDI{B}^{(1)}_{1},\cdots,\fDI{B}^{(1)}_{n^B_1},\cdots,\fDI{B}^{(m_B)}_{1},\cdots,\fDI{B}^{(m_B)}_{n^B_{m_B}}) \ .
\end{split}
\end{equation}
These permutations can be absorbed in $V$ with a redefinition of the applied transformation.
We can now introduce a compact expression for the spectral decomposition of commuting operators:
\begin{equation}
\label{eq:final_spectral_decomposition}
    \begin{split}
    \fOP{A}^{(x)}_{j_{x}} & = \sum_{a=1}^d(\Gamma^{(x)}_A)_{j_{x},a}\fOP{\Lambda}^{(x)}_{a}\,, \\
    \fOP{B}^{(y)}_{j_{y}} & = \sum_{b=1}^d(\Gamma^{(y)}_B)_{j_{y},b}\fOP{\Pi}^{(y)}_{b}\,,
    \end{split}
\end{equation}
where the rows of the matrix $\Gamma_A^{(x)}$ represent the spectral decomposition of each operator.
As we did for the matrix $M$ before, we introduce block diagonal matrices $\Gamma_{A/B}=\bigoplus_x \Gamma^{(x)}_{A/B}$. 
The spectral decomposition dictates which are the eigenvalues to be inserted in the SOS decomposition \eqref{eq:trans_sos} to ensure a saturating quantum strategy:
\begin{equation}
\label{eq:trans_sos2}
\begin{split}
    S=\qty(V_A\Gamma_A {\bf \fDI{\Lambda}}-V_B \Gamma_B{\bf \fDI{\Pi}})^\dagger
   (V_A\Gamma_A {\bf \fDI{\Lambda}}-V_B \Gamma_B{\bf \fDI{\Pi}}) \,. 
\end{split}
\end{equation}
We emphasize that $\Gamma$ depends strictly on the chosen measurement scenario.
This $\Gamma$ is the relevant object to consider in the presence of commutation relations in the final optimal strategy. Its dimension reflects the fact that it is not necessary to consider distinct sets of projectors if they are associated with commuting operators.

\paragraph{Imposing the final scenario}
In principle, we could also force from the beginning the commutation relations between different primed in order to obtain a certain final Bell scenario, and this requirement restricts in general the possible $V_{A/B}$ to be applied.
For example, the requirement $\qty[\fOP{A}_{k_1},\fOP{A}_{k_2}]=0$ translates into the additional constraint
\begin{equation}
\label{eq:constr_commutator}
\begin{split}
0 = \qty[\fOP{A}_{k_1},\fOP{A}_{k_2}] &= \sum_{s,\ell} (V_A^{-1})_{k_1s}\qty[(V_A^{-1})_{k_2\ell}]^* \,\qty[\iOP{\bf A}_{k},\qty(\iOP{\bf A}_{\ell})^\dagger] \\
&= \sum_{s,\ell,\mu,\nu} (V_A^{-1})_{k_1s}\qty[(V_A^{-1})_{k_2\ell}]^* \,M_{s\mu} \qty(M_{\ell,\nu})^*\,\qty[\iOP{\bf \Lambda}_{\mu},\iOP{\bf \Lambda}_{\nu}]\,,
\end{split}
\end{equation}
for $V_A$ to satisfy.
Above condition highlights two key dependencies of the solution. 
First, the structure of $V$ depends on the initial choice of Bell operator. 
This fact is explicit in the second line of \eqref{eq:constr_commutator}, when expressing the normal operators in terms of projectors:
different choices of $M$ may lead to distinct allowed transformations. 
Second, the possible solutions for $V$ are intrinsically linked to the commutation relations of the initial optimal strategy. 
As we will discuss in Sec.\ \ref{sec:high_dim}, certain scenarios may preclude non-trivial solutions for specific choices of $M$ and initial strategies.

\paragraph{Allowed transformations $V$} 
We now consider the most general form of the transformations $V$. For simplicity, we omit the subscripts $A$ and $B$, with the understanding that the following discussion applies separately to Alice and Bob. 
Since products of operators associated with the same inputs may appear in the Bell operator, we require that
\begin{equation}
\label{eq:block_diag_V}
    V^\dagger V = \bigoplus_{x=1}^m K^{(x)} \,,
\end{equation}
where each $K^{(x)}$ is a $n_x \times n_x$ matrix acting between operators of the same input. 
By applying the polar decomposition to $V$, we can then write:
\begin{equation}
\label{eq:gen_V}
    V=U \Sigma  \,,\qquad \Sigma=\qty(V^\dagger V)^{1/2}\,,
\end{equation}
where $U \in \text{U}(N)$ is a unitary matrix and $\Sigma$ Hermitian. 
The matrix $\Sigma$ inherits the block-diagonal structure  of $V^\dagger V$, allowing it to be expressed in terms of a direct sum of $m$ blocks $\Sigma^{(x)}$, each acting exclusively on the operators of a single input: $\Sigma = \bigoplus_x \Sigma^{(x)}$.

However, without losing generality, we could freely set $\Sigma$ to be the identity.
Indeed, since $\Sigma$ is block-diagonal and invertible, it does not alter the commutation relations \eqref{eq:constr_commutator} solved by $U$.
Consequently, the two different transformations $V_{1,A}=U_A\Sigma_A$ and $V_{2,A}=U_A$ define the same final projectors ${\bf \fOP{\Lambda}}$.
We have that
\begin{equation}
    \begin{aligned}
    {\bf \fOP{A}}&=
    V^{-1}_{1} {\bf \iOP{A}} = \Gamma_{1,A} {\bf \fOP{\Lambda}}\,, \\
    {\bf \widetilde{A''}}&= 
    V^{-1}_{2}{\bf \iOP{A}}
    = \Gamma_{2,A}  {\bf \fOP{\Lambda}}\,,
    \end{aligned}
    \quad\Rightarrow\quad
    V_{1,A}\Gamma_{1,A}=V_{2,A}\Gamma_{2,A}
\end{equation}
and then $V_{1,A}$ and $V_{2,A}$ lead to the same final SOS \eqref{eq:trans_sos2}.
Therefore, even in presence of commutation relations, the generic transformation in Eq.\ \eqref{eq:AVA} can be considered unitary without loss of generality.

\subsection{Symmetries of the derivation}
\label{sec:symmetries}
Before summarizing our discussion and presenting the method in a compact form, we address the symmetries of our approach.
Identifying symmetries that do not alter the families of obtainable Tsirelson bounds allows us to simplify computations and restrict the set of transformations under consideration.

\subsubsection{Symmetries from the choice of the initial Bell operator}
The initial Bell operator $S_0$ in \eqref{eq:SOS_AB} is defined in terms of generic normal operators, gathered in the vectors  $\mathbf{A}$ and $\mathbf{B}$ and determined by the matrix $M$.
In general, the matrix $M$ has no restrictions but different choices of it lead to the same family of Bell operators $S$.
\begin{itemize}
\item \textit{Scale invariance}: Choices of $M$ that differ only by a global scaling factor correspond to a simple rescaling of the Bell operator $S_0$, which is physically irrelevant.
In fact, the commutation constraints are not affected, and the final optimal strategies $\fOP{\bf{A}}$ and $\fOP{\bf{B}}$ are invariant.
\item \textit{Shift invariance}:  For each initial input $x$, we can perform the shift:
\begin{equation}
\label{eq:shift_M_id}
M^{(x)} \to M^{(x)} + \mathbf{c}^{(x)} \mathbf{j}^T
\end{equation}
where $\mathbf{c}^{(x)} = (c_1^{(x)}, \dots, c_{n_x}^{(x)})^T$ is an $n_x$ dimensional vector and $\mathbf{j} = (1, \dots, 1)^T$ is a $d$-dimensional vector.
By utilizing the completeness relations $\sum_{i_x} \Lambda_{i_x}^{(x)} = \sum_{i_x} \Pi_{i_x}^{(x)} = \openone$, we see that these transformations correspond to adding terms proportional to the identity to the corresponding normal operators:
\begin{equation}
A_{j_x}^{(x)} \to A_{j_x}^{(x)} + c_{j_x}^{(x)}\openone, \qquad B_{j_x}^{(x)} \to B_{j_x}^{(x)} + c_{j_x}^{(x)} \openone \ .
\end{equation}
Since the operator $S_0$ depends only on the differences $A_{j_x}^{(x)} - B_{j_x}^{(x)}$, the global Bell operator remains invariant under such shifts.
Moreover, since a term proportional to the identity does not affect the commutativity constraints \eqref{eq:commutator_condition}, the set of allowed unitary transformations $U_{A/B}$ remains the same, and the transformed operators share the same projectors.
Finally, for a fixed transformation $U$, the shift \eqref{eq:shift_M_id} does not change the final Bell operators.
To see this, consider the cases without and with the shift, applying the rotation only to Bob's side for simplicity:
\begin{equation}
\begin{aligned}
U_B^{-1}\iOP{\bf{B}}&= \Gamma_B \fOP{\Pi} \\
U_B^{-1}(\iOP{\bf{B}}+\mathbf{c}\mathbf{1})&= \Gamma_B' \fOP{\Pi}
\end{aligned}
\qquad \implies \qquad \Gamma_B \fOP{\Pi}=\Gamma_B'\fOP{\Pi}-U_B^{-1}\mathbf{c}\mathbf{1}
\end{equation}
where we used the fact that the set of projectors of the transformed operators remains the same.
The corresponding final Bell operators are:
\begin{equation}
\begin{aligned}
S&=(\Gamma_A \mathbf{\Lambda}'-U_B \Gamma_B \mathbf{\Pi}')^\dagger (\Gamma_A \mathbf{\Lambda}'-U_B \Gamma_B \mathbf{\Pi}') \\
S'& =(\Gamma_A \mathbf{\Lambda}'+\mathbf{c}\mathbf{1}-U_B \Gamma_B' \mathbf{\Pi}')^\dagger (\Gamma_A \mathbf{\Lambda}'+\mathbf{c}\mathbf{1}-U_B \Gamma_B' \mathbf{\Pi}')=S \ .
\end{aligned}
\end{equation}
This implies that it is sufficient to consider only initial traceless normal operators.

\item \textit{Unitary invariance}: Since the SOS construction $\sum_i P_i^\dagger P_i$ involves the product $(M^{(x)})^\dagger M^{(x)}$, there is the freedom to multiply $M^{(x)}$ from the left by a unitary matrix. In other words, by applying the singular value decomposition (SVD):
\begin{equation}
\label{eq:decomp_M}
    M^{(x)} = Y^{(x)} D^{(x)} W^{(x)}\,,
\end{equation}
where $D^{(x)}$ is a positive semi-definite diagonal matrix, and $Y^{(x)}$ and $W^{(x)}$ are unitary matrices, we can set $Y^{(x)} = \openone$ without loss of generality. This reduces the decomposition to:
\begin{equation}
    M^{(x)} = D^{(x)} W^{(x)} \,.
\end{equation}
In terms of allowed transformations, $Y^{(x)}$ can be absorbed into the unitary matrix $U$, which does not change the set of possible final operators.
\end{itemize}

We will further explore the utility of these symmetries in Sec.\ \ref{sec:qubit_strate} to simplify our computations.
Their remarkable property is that they reduce the set of initial operators $S_0$ that need to be considered, without restricting the set of final $S$ operators that can be obtained.

\subsubsection{Transforming one party is sufficient}
Previously, we arrived at the expression
\begin{equation}
S=\mathbf{G}_0^\dagger \mathbf{G}_0\,, \qquad {\bf G}_0=U_A \Gamma_A \boldsymbol{\Lambda}'-U_B \Gamma_B \boldsymbol{\Pi}'\,
\end{equation}
for the final Bell operator.
Note that $S$ remains unchanged if we apply the unitary $U_A^\dagger$ to ${\bf G}_0$:
\begin{equation}
    S = {\bf G}_1^\dagger {\bf G}_1\,, \qquad {\bf G}_1= U_A^\dagger \mathbf{G}_0 =\Gamma_A{\bf\Lambda}'-U_A^\dagger U_B\Gamma_B{\bf \Pi}' \,.
\end{equation}
This expression implies that it is sufficient to consider transformations on only one of the sides. Indeed, as discussed, the matrix $\Gamma_A$ is analogous to the matrix $M$ described earlier.
By considering the saturating strategy $\Gamma_A \widetilde{\bf \Lambda'}$ for ${\bf G}_0$ as the initial strategy for obtaining ${\bf G}_1$, $\Gamma_B$ becomes exactly the spectral decomposition that yields Bob's operator after the transformation $U_A^\dagger U_B$ on Bob's side. 
To see this explicitly, note that the relations between the spectral decompositions of the initial operators and the transformed operators are (cf. \eqref{eq:opti_strate_U}):
\begin{equation}
\begin{aligned}
    M \iOP{\bf{\Lambda}}&=U_A\Gamma_A \fOP{\bf{\Lambda}}\,, \\
    M \iOP{\bf{\Pi}}&= M(\iOP{\bf{\Lambda}}^T) = U_B\Gamma_B \fOP{\bf{\Pi}}\,,
\end{aligned}
\end{equation}
(where the transpose acts on the Hilbert space of the operators). 
This also implies:
\begin{equation}
\Gamma_B \fOP{\bf{\Pi}}=U_B^\dagger M(\iOP{\bf\Lambda}^T)=(U_A^\dagger U_B)^{-1} \Gamma_A (\fOP{\bf\Lambda})^T\,.
\end{equation}
We can then obtain $S$ by starting from an initial strategy with $\Gamma_A \bf \widetilde{\Lambda}'$ instead of $M \bf \widetilde{\Lambda}$, and applying $U_A^\dagger U_B$ on Bob's side.
Therefore, without loss of generality, we can set $U_A=\openone$.

\subsection{Final Bell operators: summary}
Summarizing the previous discussion and removing irrelevant superscripts in the final expression, the general SOS we derived is 
\begin{equation}
\label{eq:final_bell}
    S=\mathbf{G}^\dagger \mathbf{G}\geq0\,,\qquad \mathbf{G}=DW{\bf \iDI{\Lambda}}-U \Gamma {\bf \iDI{\Pi}}\,.
\end{equation}
In this expression we can freely choose the matrices $D$ and $W$ and a generic optimal strategy $\iOP{\mathbf{A}}=DW\widetilde{\bf\Lambda}$ on Alice's side. 
The unitary matrix $U$ must be then chosen in order to satisfy the normality constraint \eqref{eq:commutator_condition}.
The optimal Bob's strategy is obtained by applying $U^{-1}$ to the transpose of Alice's strategy and the matrix $\Gamma$ is determined by the spectral decomposition of the normal operators obtained,
as shown in \eqref{eq:final_spectral_decomposition}.
The commutation properties of these operators dictate the number of inputs on Bob's side.

Because Alice's strategy can be expressed in terms of traceless operators (as shown in the previous section), Bob's final strategy is also traceless, as it is defined as a linear combination of Alice's operators.

\subsubsection{Local bounds}
While $\expval{S}=0$ defines a quantum bound, the local bound, namely, the minimum expectation value achievable by local hidden variable theories, will generally be strictly greater than zero. 
For LHV theories, the set of correlations forms a convex polytope; thus, any local probabilistic behavior can be expressed as a convex combination of its vertices, which correspond to a finite number of deterministic strategies \cite{Fine1982,Brunner2014}. 
Because the expectation value is a linear functional of the local correlations, minimizing it over the entire polytope is equivalent to performing the minimization over its vertices.

We represent the deterministic vertices using vectors of classical binary variables, $\boldsymbol{\lambda}$ and $\boldsymbol{\pi}$. Their elements, $\lambda^{(x)}_a, \pi^{(y)}_b \in \{0,1\}$, define a deterministic assignment of outputs for each respective input. 
To ensure completeness, where exactly one outcome occurs per measurement setting, we impose the constraints:
\begin{equation}
    \sum_{a=1}^d \lambda^{(x)}_a = \sum_{b=1}^d \pi^{(y)}_b=1 \ .
\end{equation}
Subject to these boolean and normalization constraints, the classical variables algebraically mimic the idempotence ($P^2 = P$) and mutual orthogonality ($P_i P_j = 0$ for $i \neq j$) of the quantum projector vectors $\mathbf{\Lambda}$ and $\mathbf{\Pi}$. 
This structural parallel allows us to compute the local bound $S_{\rm LHV}$ directly from the sum-of-squares decomposition without expansion. 
The problem then reduces to minimizing the squared norm of the vector $\mathbf{g}$ over all valid deterministic assignments:
\begin{equation}
\label{eq:LHV_bound}
\mathcal{S}_{\rm LHV} = \min_{\boldsymbol{\lambda},\boldsymbol{\pi}} 
\|\mathbf{g}\|^2\,,\qquad \mathbf{g} = DW\boldsymbol{\lambda}-U\Gamma \boldsymbol{\pi} \ ,
\end{equation}
which gives the local bound for a given choice of initial optimal strategy and unitary transformation.

We are now ready to summarize the method discussed so far.

\begin{recipebox}{Recipe to obtain Bell operators with a known Tsirelson bound}
\begin{enumerate}
    \item \textit{Fix Alice's scenario:} Fix a scenario on Alice's side with $m_A$ inputs and $d$ outputs and a corresponding Hilbert space $\mathcal H_A$.
    \item \textit{Choose Alice's optimal strategy:} Choose $m_A$  sets of $d$ orthogonal projectors $\{\widetilde\Lambda^{(x)}_a\}$ 
    ($\widetilde\Lambda^{(x)}_a\widetilde\Lambda^{(x)}_b=\delta_{ab}\widetilde\Lambda^{(x)}_a$ and $\sum_a^{(x)}\widetilde\Lambda^{(x)}_a=\openone$ for each $x$)
    acting on $\mathcal H_A$. The index $x=1,\dots,m_A$ defines the input, and the subscript $a=1,\dots,d$ identifies different operators of the same input. Collect the $d\cdot m_A$ projectors in a vector ${\bf \widetilde\Lambda}$ with elements $\widetilde\Lambda^{(x)}_a$.

    \item \textit{Choose Alice's normal operators:}
    For each $x$, choose a $n^A_x \times d$ matrix $M^{(x)}$. Define $M=\bigoplus_{x} M^{(x)}$ and combine the projectors into the normal operators:
    \begin{equation}
    \begin{aligned}
    \widetilde{A}_{i_x}^{(x)}=(M \widetilde{\bf\Lambda})^{(x)}_{i_x} \,
    \qquad i_x=1,\dots , n^A_x. 
    \end{aligned}
    \end{equation} 
    It is sufficient to consider $M^{(x)}$ such that $M^{(x)}=D^{(x)}W^{(x)}$ with $D^{(x)}$ real and rectangular diagonal matrix, $W^{(x)}$ unitary, and such that $\Tr[M^\dagger M] = 1$ and the operators $\widetilde{A}_{i_x}^{(x)}$ are traceless. 
    Define $N=\sum_x n^A_x$ as the total number of normal operators 
    $\{\widetilde{A}_{i_x}^{(x)}\}$.
    \item \textit{Choose Bob's transformation:} Define $\widetilde{B}^{(x)}_{i_x}={(\widetilde{A}^{(x)}_{i_x})}^T$ and gather all the $\widetilde{B}^{(x)}_{i_x}$ in a single vector $\mathbf{\iOP{B}}$.
    Choose a unitary transformation, $U\in {\rm U}(N)$ which satisfies the following constraint:
    \begin{equation}
        \label{eq:recipe_constr}
        \qty[{\widetilde{B'}}_{k},
        {({\widetilde{B'}}_{k})}^\dag]= 0 \ \qquad \forall k \ ,
        \end{equation}
        where
    \begin{equation}
    \label{eq:recipe_op}
    \begin{aligned}
        {\widetilde{B'}}_k=(U^{-1} \widetilde{\bf B})_{k} \ . 
    \end{aligned}
    \end{equation}
    As before, gather $  {\widetilde{B'}}_k$ into a vector $\mathbf{\fOP{B}}$.
     \item \textit{Define Bob's scenario:} Identify the operators ${\widetilde{B'}}_k$ which commute and gather them into distinct sets. The number of such sets is the number $m_B$ of Bob's inputs, each one with $d$ outcomes. The number of commuting operators
     in each set is denoted by $n^B_y$.
     Rename the elements $\widetilde{B'}_k$ of each set $y$ and call them 
     $ {\widetilde{B'}}_{j_y}^{(y)}$
     with $j_y=1,\cdots,n^B_y$ and $\sum_{y=1}^{m_B}n^B_y=N$. 
     If operators of the same $y$ are not close in the vector $\mathbf{\fOP{B}}$, consider a permuted version of $U$ to make this happen.
    \item 
    \textit{Determine Bob's spectral decomposition:}
    Determine the matrix $\Gamma=\bigoplus_y\Gamma^{(y)}$  by the following procedure. Due to condition 
    \eqref{eq:recipe_constr}, the operators in \eqref{eq:recipe_op} are normal and those associated to the same input commute. Therefore, at fixed input, they could be simultaneously diagonalized.
    The matrices $\Gamma^{(y)}$ are determined by the following spectral decomposition:
    \begin{equation}
    \label{eq:final_strategy}
    {\widetilde{B'}}_{j_y}^{(y)}= \sum_{b=1}^d\Gamma^{(y)}_{j_yb}\widetilde{\Pi}^{(y)}_{b}\,,\qquad j_y=1,\cdots,n^B_y
    \end{equation}
    with $\widetilde{\Pi}^{(y)}_b$ projectors. 
    \item \textit{Bell operator:}
    Define ${\bf \Lambda}$ as the vector with elements $\Lambda^{(x)}_a$, and ${\bf \Pi}$ the vector with elements $\Pi^{(y)}_b$, where $\Lambda^{(x)}_a$ and $\Pi^{(y)}_b$ are symbolic projectors. The final SOS Bell operator is 
    \begin{equation}
    \label{eq:final_S}
    S = \mathbf{G}^\dag \mathbf{G}\,,\qquad \mathbf{G}=DW{\bf \Lambda}-U\Gamma{\bf \Pi}\,.
    \end{equation}
    The quantum bound $\ev{S}= 0$ is saturated by the following strategy:
\begin{equation}
    \ket{\Phi^+}=\frac{1}{\sqrt{d}}\sum_{i=1}^d \ket{i}\ket{i}\,,
\quad\Lambda^{(x)}_{a}=\widetilde{\Lambda}^{(x)}_{a}\,,
\quad\Pi^{(y)}_{b}=\widetilde{\Pi}^{(y)}_{b}
\end{equation} 
with $\widetilde{\Lambda}^{(x)}_{a}$ the initially chosen orthogonal projectors
and $\widetilde{\Pi}^{(y)}_{b}$ determined by Eq.\ \eqref{eq:final_strategy}. 
    \item \textit{Local bound:}
    The local bound is obtained by minimizing the norm of the vector $\bf g$ 
    \begin{equation}
        \mathcal{S}_{\rm LHV}\geq \beta=
        \min_{\{\boldsymbol{\lambda},\boldsymbol{\pi}\}} \| {\bf g}\|^2\,,\qquad
        {\bf g} = DW\boldsymbol{\lambda}-U\Gamma \boldsymbol{\pi}   
    \end{equation}
        where the elements $\lambda^{(x)}_a, \pi^{(y)}_b \in \{0,1\}$ of the vectors $\boldsymbol{\lambda}$ and $\boldsymbol{\pi}$ 
        satisfy $\sum_a\lambda^{(x)}_a=\sum_b\pi^{(y)}_b=1$, $\forall x,y$
        and therefore define a deterministic assignment of outputs for each respective input.
\end{enumerate}
\end{recipebox}

In the following, we apply this method to both qubit and qudit strategies.
In both cases we will provide analytical solutions for the constraint in Eq.\ \eqref{eq:recipe_constr}, reproducing new results and obtaining new Bell operators with the associated Tsirelson bounds and saturating strategies.

\section{Qubit inequalities}
\label{sec:qubit_strate}
We now proceed to a detailed analysis of the Tsirelson bounds that can be saturated by generic qubit strategies acting on a maximally entangled state, in a scenario in which Alice can perform $m$ different measurements. 
These strategies have two outputs for each input so that we require $d=2$.
Following the discussion of Sec.\ \ref{sec:symmetries}, we can consider traceless normal operators for Alice's initial optimal strategy.
We note that any normal traceless $2\times 2$ operator can be written as a phase term times a rescaled traceless unitary operator, $\iDI{A}^{(x)}_{j_x} = \exp[i \theta_{j_x}]k_{j_x}\iDI{\mathcal{A}}^{(x)}_{j_x}$.
This choice aligns us with the existing literature, where qubit Bell inequalities are typically expressed in terms of unitary operators.
We then write the initial Bell operator as
\begin{equation}
    S_0 = (\iDI{\bf \Lambda} - \iDI{\bf \Pi})^\dagger M^\dagger M (\iDI{\bf \Lambda} - \iDI{\bf \Pi})= (\iDI{\bf A} - \iDI{\bf B})^\dagger (\iDI{\bf A} - \iDI{\bf B}) = \frac{1}{2}(\iDI{\boldsymbol{\mathcal A}} - \iDI{\boldsymbol{\mathcal B}}) K^\dagger K (\iDI{\boldsymbol{\mathcal A}} - \iDI{\boldsymbol{\mathcal B}})\,,
    \end{equation}
where the vectors $\boldsymbol{\mathcal A},\boldsymbol{\mathcal B}$ group the unitary operators, and the $N\times N$ diagonal matrix $K$ groups the coefficients $\sqrt2k_{j_x}$.
Since $M$ is defined up to a global constant, then also $K$ is; we therefore impose $\Tr[K^2] = 1$ ($\sum_j k_j^2 = 1/2$).

Note that the phase terms can be removed thanks to the unitary invariance described in Sec.\ \ref{sec:symmetries}, so Alice's initial optimal strategy, saturating the Tsirelson bound of $S_0$, is
\begin{equation}
    \iOP{\mathcal A}^{(x)}_{j_x}={\vec n }^A_{j_x}\cdot\vec\sigma\,,\qquad |\vec n_{j_x}^A|=1
\end{equation}
a generic traceless unitary observable. 
Here we label three dimensional Euclidean vectors with an arrow, $\vec n$, for differentiating from the vectors of operators defined previously.
Bob's initial optimal strategy is the transpose of Alice's in the canonical basis
$\iOP{{\mathcal B}}^{(x)}_{j_x}={\vec n}^B_{j_x}\cdot\vec\sigma$ with ${\vec n}^B_{j_x}=(n^A_{1,j_x},-n^A_{2,j_x},n^A_{3,j_x})$. 

We seek the general form of an $N \times N$ unitary matrix $U$ satisfying the condition (cf.\ \eqref{eq:commutator_condition}):
\begin{equation}    
\label{eq:qubit_m_input} 
    \qty[\fOP{B}_j,\fOP{B}_j^\dagger]= \sum_{p, \ell = 1}^N U^{-1}_{jp} (U^{-1}_{j\ell})^* \qty[ \iOP{\bf B}_{p}, (\iOP{\bf B}_{\ell})^\dagger ] = \sum_{p, \ell = 1}^N U^{-1}_{jp} (U^{-1}_{j\ell})^* k_p k_\ell\qty[ \iOP{\boldsymbol{\mathcal{B}}}_{p}, (\iOP{\boldsymbol{\mathcal{B}}}_{\ell})^\dagger ] =0   \ .
\end{equation}
This equation requires the transformed operators, $\fOP{B}_j = \sum_k U^{-1}_{jk} \iOP{\bf B}_k$, to be normal. 
Since we started with initial traceless operators and performed linear combinations, the resulting operators will also be traceless. 
As already noted, any normal traceless $2\times 2$ operator can be written as a phase term times an Hermitian traceless matrix:
$\fOP{B}_j= e^{i\theta_j}H_j = \sum_{k=1}^N U^{-1}_{jk} \iOP{\bf B}_k$, with $H_j = H_j^\dagger$.
Multiplying both sides by $e^{-i\theta_j}$ and requiring the right-hand side to be Hermitian, we obtain (note that $\iOP{\bf B}_k$ is Hermitian):
\begin{equation}
\label{eq:qubit_norm_cond}
        \sum_{k=1}^N \qty(e^{-i \theta_j}U^{-1}_{jk} - e^{i \theta_j}(U^{-1}_{jk})^*) \iOP{\bf B}_k = 0 \ .
\end{equation}
For every choice of initial strategy $\iOP{\bf B}_k$ a solution of the previous equation is found by requiring that $e^{-i \theta_j}U^{-1}_{jk}$ are the components of a real matrix, $R$. 
This is how to say that real combinations of Hermitian operators are always Hermitian.
If the initial strategy consists of linearly independent operators $\iOP{\bf B}_k$ (e.g., the three Pauli matrices) $e^{-i \theta_j}U^{-1}_{jk}\in \mathbb{R}$ is \textit{the only possible solution}. 
More general cases are discussed in Appendix \ref{sec:ind_qubit}.
We note that commuting normal $2\times 2$ matrices are proportional, implying that the initial optimal strategy with more than one operator per input always has some dependencies.
Such dependencies can be reduced by considering a matrix $M$ which reduces the number of normal operators for each measurement in the initial strategy to one.

As a last step, we need to require the operator $U^{-1}_{jk}=e^{i \theta_j}R_{jk}$ to be the components of an unitary matrix. 
This implies that $U$ must be a diagonal phase matrix times an orthogonal one:
    \begin{equation}
        U= \Phi O\,,
    \end{equation}
where $\Phi=\text{diag}(e^{-i \theta_1}, \dots , e^{-i \theta_N})$ and $O \in {\rm SO}(N)$.

Since the matrix $\Phi$ does not mix different inputs, as we discussed in \ref{sec:scenarios} for  the matrix $\Sigma$, we can set $\Phi=\openone_m$ since its contribution disappears when performing the spectral decomposition of the transformed operators.
The new operators after the transformation can be written as
\begin{equation}
    \widetilde B'_k=\sum_{k}O^{-1}_{kj}\widetilde{\bf B}_j\,,
\end{equation}
which can be expressed in terms of unitary operators as
\begin{equation}
    \widetilde{\mathcal B}'_x=\frac{1}{|\vec u^B_x|}\vec u^B_x\cdot\vec\sigma\,,\qquad
        \vec u^B_x=\sqrt{2}\sum_{x'}O^{-1}_{xx'}\,k_{x'}\vec n^B_{x'}\,.
\end{equation}
The original Bell operator, $S_0$, is then transformed into 
\begin{equation}
         S = \frac{1}{2}\mathbf{G'}^\dag \mathbf{G'}\,,\qquad 
    \mathbf{G'}=K\boldsymbol{\mathcal A}'-O\Gamma\boldsymbol{\mathcal B}'\,,
\end{equation}
where $\Gamma = {\rm diag}(| {\vec u^{B}_1}|, | {\vec u^{B}_2}|,\cdots,   | {\vec u^{B}_N}|)$. 
By expanding the product we get
\begin{align} 
         S &=\frac{1}{2}(\boldsymbol{\mathcal A}')^\dag K^2\boldsymbol{\mathcal A}'+
         \frac{1}{2}\boldsymbol{\mathcal B}'^\dag\Gamma^2\boldsymbol{\mathcal B}'-
         (\boldsymbol{\mathcal A}')^\dag K O\Gamma\boldsymbol{\mathcal B}' \\
         &= \openone - (\boldsymbol{\mathcal A}')^\dag K O\Gamma\boldsymbol{\mathcal B}'\,,
\end{align}
where we computed $(\boldsymbol{\mathcal A}')^\dag K^2\boldsymbol{\mathcal A}'=\Tr[K^2] \openone = \openone $, and $(\boldsymbol{\mathcal B}')^\dag\Gamma^2\boldsymbol{\mathcal B}' = \Tr[\Gamma^2]\openone = \sum_x |\vec u^{B}_x|^2 \openone = \sum_x 2k^2_x|\vec n^{B}_x|^2 \openone = \Tr[K^2]\openone$.

\paragraph{Summary for qubit Tsirelson bounds}
Here we summarize how our method derives Tsirelson bounds saturated by qubit strategies.
\begin{enumerate}
    \item Choose $N$ 3-dimensional unit vectors $\vec n^A_x$, $x=1,\cdots,N$ 
    ($|\vec n_x^A|=1$). 
    Define the unit vectors ${\vec n}^B_x=(n^A_{1,x},-n^A_{2,x},n^A_{3,x})$.
    These vectors correspond to $N$ unitary operators defining Alice's strategy. 
    The number of sets of parallel vectors, $m$, defines the number of inputs of Alice, each one with 2 outcomes.
    \item Choose a non-negative $N\times N$ diagonal matrix $K$ with elements $K_{xy}=\sqrt{2}k_x\delta_{xy}$ with the normalization condition $\Tr[K^2] = \sum_x 2k_x^2=1$.
    \item Choose an orthogonal matrix with unit determinant, $O \in {\rm SO}(N)$. If the vectors $\vec n^A_x$ are linearly dependent, you can also choose some non-orthogonal matrices characterized in Appendix \ref{sec:ind_qubit}.
    \item Define the vectors 
    \begin{equation}
        \vec u^{B}_x=\sqrt{2}\sum_{y=1}^N (O^{-1})_{xy}\,k_y\,\vec n^{B}_{y} \ . 
    \end{equation}
    \item Define ${\boldsymbol{\mathcal A}}$ as the vector with elements $\mathcal A_x$, and $\boldsymbol{\mathcal B}$ the vector with elements $\mathcal B_x$, where $\mathcal A_x$ and $\mathcal B_x$ are symbolic unitary Hermitian operators. 
    The Bell operator is given by
    \begin{equation}
    \label{eq:final_S_qubit}
    \begin{aligned}
    \mathbf{G'}&=K{\boldsymbol{\mathcal A}}-O\Gamma{\boldsymbol{\mathcal B}}\,,
    \\
    S &= \frac{1}{2}\mathbf{G'}^\dag \mathbf{G'}
    =\openone-{\boldsymbol{\mathcal A}}^\dag KO\Gamma {\boldsymbol{\mathcal B}}
    \,,\qquad 
    \end{aligned}
    \end{equation}   
    where  
    \begin{equation}
        \Gamma={\rm diag}(| {\vec u^{B}_1}|,
        | {\vec u^{B}_2}|,\cdots,   | {\vec u^{B}_N}|)\,.
    \end{equation}
    \item The quantum bound
    $\ev{S}\geq 0$ is saturated by the following strategy:
\begin{equation}
\label{eq:optimal_strategy_qubit}
    \ket*{\Phi^+}=\frac{\ket{00}+\ket{11}}{\sqrt{2}}\,,
\quad
\iOP{\mathcal{A}}_x=\vec n^A_x\cdot\vec\sigma\,,\qquad
\iOP{\mathcal B}_x=\frac{1}{|\vec u^B_x|}\vec u^B_x\cdot\vec\sigma\,.
\end{equation} 
    \item 
    The classical bound is obtained by minimizing the following function:
    \begin{equation}
        \mathcal{S}_{\rm LHV}\geq \beta=
        \min_{\{\boldsymbol{a},\boldsymbol{b}\}} g^2({\bf a},{\bf b})\,,\qquad
        g^2({\bf a},{\bf b}) = 1-\,{\bf a}^T KO\Gamma {\bf b}   
    \end{equation}
where the elements $a_x, b_x \in \{-1,+1\}$ of the vectors $\mathbf{a}$ and $\mathbf{b}$ define a deterministic assignment for each measurement.
        We note that $g^2({\bf a},{\bf b})$ has $2^{2N}$ discrete
        possible values.
\end{enumerate}

\subsection{Minimal scenario: two inputs per party}
We now restrict our attention to the case $N=m$, namely, the case in which there is only one normal operator per input. 
As noted above, this case is relevant for qubits, since commuting traceless operators (such as those associated with the same input) are essentially the same operator, up to an overall scalar factor.
In the minimal scenario $(2,2,2)$, both parties have two inputs.
Alice's strategy is defined by two unit vectors $\vec n^A_1$ and  $\vec n^A_2$
and the corresponding Bob's initial strategy is ${\vec n}^B_x=(n^A_{1,x},-n^A_{2,x},n^A_{3,x})$.
The two numbers $k_{1,2}$ that satisfy $k^2_1+k^2_2=1/2$ can be chosen as
\begin{equation}
    k_1=\frac{\cos\gamma}{\sqrt{2}}\,,\qquad k_2=\frac{\sin\gamma}{\sqrt{2}}\,,\qquad 0\leq\gamma\leq\frac\pi2 \ .
\end{equation}
The general orthogonal matrix $O$ can be written as 
\begin{equation}
\label{eq:rotation_O}
        O =
    \begin{pmatrix}
        \cos\theta & \sin\theta
        \\
        -\sin\theta & \cos\theta
    \end{pmatrix}\,,
\end{equation}
from which the vectors $\vec u^{B}_x=\sqrt{2}\sum_{y=1}^2 (O^{-1})_{xy}\,k_y\vec n^{B}_{y}$ can be derived:
\begin{equation}
    \begin{aligned}
        \vec u^B_1&=\cos\theta \cos\gamma \,\vec n^B_1 
        - \sin\theta\sin\gamma \,\vec n^B_2 \ , \\
        \vec u^B_2&=\sin\theta \cos\gamma\,\vec n^B_1 
        + \cos\theta \sin\gamma\,\vec n^B_2 \ .
    \end{aligned}
\end{equation}
The vector ${\bf G'}$ and 
the final Bell operator are then given by
\begin{equation}
    \begin{aligned}
    \mathbf{G'} &= 
    \begin{pmatrix}
    \cos\gamma \mathcal{A}_1 \\
    \sin\gamma \mathcal{A}_2
    \end{pmatrix} -  \begin{pmatrix}
    \cos\theta & \sin\theta \\
    -\sin\theta & \cos \theta
    \end{pmatrix}
    \begin{pmatrix}
        |\vec u^B_1| & 0 \\
        0 & |\vec u^B_2|
    \end{pmatrix}
    \begin{pmatrix}
    \mathcal{B}_1 \\
    \mathcal{B}_2
    \end{pmatrix} \label{eq:G_final_trans}\,,
    \\
    S &= \openone
    -\cos\theta\Bigl[\cos\gamma |\vec u^B_1| \mathcal{A}_1\mathcal{B}_1+\sin\gamma |\vec u^B_2| \mathcal{A}_2\mathcal{B}_2\Bigr]-\sin\theta \Bigl[\cos\gamma |\vec u^B_2| \mathcal{A}_1\mathcal{B}_2-
     \sin\gamma |\vec u^B_1| \mathcal{A}_2\mathcal{B}_1\Bigr] \ .
    \end{aligned}
\end{equation}
The Tsirelson bound $\ev{S}\geq0$ is saturated by measuring the unitary observables
\begin{equation}
    \begin{aligned}
        \widetilde{\mathcal{A}}_1&=\vec n^A_1\cdot\vec\sigma\,,\qquad & \widetilde{\mathcal{B}}_1&=\frac{1}{|\vec u^B_1|}
        \vec u^B_1\cdot\vec\sigma \ ,
        \\
        \widetilde{\mathcal{A}}_2&=\vec n^A_2\cdot\vec\sigma\,,\qquad & \widetilde{\mathcal{B}}_2&=\frac{1}{|\vec u^B_2|} 
        \vec u^B_2\cdot\vec\sigma \ .
    \end{aligned}
\end{equation}
The local bound can be explicitly evaluated as
\begin{equation}
    \mathcal{S}_{\rm LHV}\geq \beta =\min\{\beta_1,\beta_2\}\,,
\end{equation}
where
\begin{equation}
\label{eq:local_bound_qubit}
\begin{aligned}
    \beta_1&=1
    -\cos\gamma \abs{\cos\theta |\vec u^B_1| +\sin\theta \abs{\vec u^B_2}}
    -\sin\gamma\abs{ \sin\theta |\vec u^B_1| -
    \cos\theta |\vec u^B_2|}\,,
\\
    \beta_2&=1
    -\cos\gamma \abs{\cos\theta |\vec u^B_1| -\sin\theta \abs{\vec u^B_2}}
    -\sin\gamma\abs{ \sin\theta |\vec u^B_1| +
    \cos\theta |\vec u^B_2|}.
\end{aligned}
\end{equation}

We note that we can always choose Alice's strategy as $\vec n^A_1=(0,0,1)$ and $\vec n^A_2=(\sin \alpha,0,\cos \alpha)$.
This simply corresponds to fixing a global reference system for final measurement eigenvectors and the state, which remains maximally entangled. 

\begin{examplebox}{A family of qubit inequalities}
We consider the family of Bell operators introduced in \cite{Barizien_2024}:
\begin{equation}
    I = \frac{2}{\sin(b_2-b_1)}\qty[ \sin(b_2)\mathcal{A}_1\mathcal{B}_1 + \frac{\sin(b_2-a_2)}{\sin^2(a_2) F}\mathcal{A}_2\mathcal{B}_1 + \frac{\sin(a_2-b_1)}{\sin^2(a_2) F}\mathcal{A}_2\mathcal{B}_2 - \sin(b_1)\mathcal{A}_1\mathcal{B}_2 ]
\end{equation}
with the quantum bound
\begin{equation}
    \mathcal{I}_{\mathcal{Q}} = \frac{2\sin(a_2)\sin(a_2-b_1-b_2)}{\sin(a_2-b_1)\sin(a_2-b_2)}\,.
\end{equation}
The bound above has the notable property of self-testing all measurement settings that can be certified by the maximally entangled state (the singlet), except for configurations in which Alice and Bob share a common measurement setting.
The optimal strategy is
\begin{equation}
\begin{split}
    \mathcal{A}_1 &= \sigma_z\,, \quad \mathcal{A}_2 = \cos(a_2)\sigma_z + \sin(a_2)\sigma_x\,, \\
    \mathcal{B}_1 &= \cos(b_1)\sigma_z + \sin(b_1)\sigma_x\,, \quad \mathcal{B}_2 = \cos(b_2)\sigma_z + \sin(b_2)\sigma_x\,.
\end{split}
\end{equation}
The related SOS in terms of unitary operators is
\begin{align}
    \mathcal{I}_{\mathcal{Q}}\openone - I &= N_1^\dagger N_1 + N_2^\dagger N_2\,, \label{eq:SOS_Bari}\\
    N_1 &= \mathcal{A}_1 - \frac{\sin(b_2)\mathcal{B}_1-\sin(b_1)\mathcal{B}_2}{\sin(b_2-b_1)}\,,\\
    N_2 &= \frac{1}{\sin(a_2)\sqrt{F}}\qty[ \mathcal{A}_2 - \frac{\sin(b_2-a_2)\mathcal{B}_1 - \sin(b_1-a_2)\mathcal{B}_2}{\sin(b_2-b_1)} ]\,,\\
    F &= \qty[\cot(a_2) - \cot(b_2)]\qty[\cot(b_1) - \cot(a_2)]\,.
\end{align}
In particular, the previous SOS holds only if $F>0$, i.e. if $b_1 < a_2 < b_2 $.

For obtaining this Tsirelson bound with our method, we normalize \eqref{eq:SOS_Bari} to write it the form \eqref{eq:final_S_qubit}:
\begin{align}
    \openone - I/\mathcal{I}_{\mathcal{Q}} &= \frac{1}{2}\left((N'_1)^\dagger N'_1 + (N'_2)^\dagger N'_2 \right)\stackrel{!}{=}\openone-{\boldsymbol{\mathcal A}}^\dag KO\Gamma {\boldsymbol{\mathcal B}}\, 
\end{align}
where we defined $ N'_i = \sqrt{2}N_i / \sqrt{\mathcal{I}_{\mathcal{Q}}}$.
The matrix $K$ is fixed by the coefficients in front of the operators $\mathcal A_1$ and $\mathcal A_2$ to be:
\begin{equation}
    K = \frac{\sqrt{2}}{\sqrt{\mathcal{I}_{\mathcal Q}}}\begin{pmatrix}
        1 & 0 \\
        0 & \frac{1}{\sin(a_2)\sqrt{F}}
    \end{pmatrix}\,,
\end{equation}
satisfying $\Tr[K^2]=1$, to retrieve the format of \eqref{eq:final_S_qubit}.
The transformation on Bob's side is instead fixed by the requirement
\begin{equation}
    O\Gamma = \frac{\sqrt{2}}{\sqrt{\mathcal{I}_{\mathcal Q}}}\begin{pmatrix}
        \frac{\sin(b_2)}{\sin(b_2-b_1)} & -\frac{\sin(b_1)}{\sin(b_2-b_1)} \\
        \frac{1}{\sin(a_2)\sqrt{F}}\frac{\sin(b_2-a_2)}{\sin(b_2-b_1)} & -\frac{1}{\sin(a_2)\sqrt{F}}\frac{\sin(b_1-a_2)}{\sin(b_2-b_1)}
    \end{pmatrix}\,,
\end{equation}
with $O$ a rotation matrix as in \eqref{eq:rotation_O}.
A possible solution is
\begin{align}
    |{\vec u^B_1}| &= \sqrt{\frac{\sin(a_2 - b_2) \sin(b_1 - b_2)}{\sin(b_2) \sin(b_1 + b_2 - a_2)}}
  \frac{\sin(b_2)}{\sin(b_2 - b_1)} \ ,\\
   |{\vec u^B_2}| &=\sqrt{ \frac{\sin(a_2 - b_1) \sin(b_1 - b_2) (\cot(a_2) - \cot(b_2))^2}{
\sin(b_1)\sin(a_2 - b_1 - b_2)}}\,, \\
   \theta &= \arctan\left(\frac{\cot (b_2)-\cot (a_2)}{\sqrt{F}}\right) \ .
\end{align}

Finally, we note that if $a_2=\pi/2$, $b_1 = \alpha$ and $b_2 = b_1 - \pi/2$, \eqref{eq:SOS_Bari} reduces to the SOS of the operator 
\begin{equation}
    \sqrt{2}(2\sqrt{2} \openone - \cos t I_{\rm CHSH} - \sin t I'_{\rm CHSH})\,,
\end{equation}
with $t = \alpha - \pi/4$ and 
\begin{align}
    I_{\rm CHSH} &= \mathcal{A}_1\mathcal{B}_1 + \mathcal{A}_1\mathcal{B}_2 + \mathcal{A}_2\mathcal{B}_1 - \mathcal{A}_2\mathcal{B}_2 \\
    I'_{\rm CHSH} &= -\mathcal{A}_1\mathcal{B}_1 + \mathcal{A}_1\mathcal{B}_2 + \mathcal{A}_2\mathcal{B}_1 + \mathcal{A}_2\mathcal{B}_2
\end{align}
introduced in \cite{Christensen2015} and representing a trade-off between two CHSH operators.

Moreover, if we choose $a_2=\delta + \pi/2$, $b_1=\pi/2$ and $b_2=-\delta$, and we divide \eqref{eq:SOS_Bari} by $2\tan(\delta)$, we retrieve the SOS of the operator
\begin{equation}
\label{eq:woolt_sos}
    \mathcal{I}'_\mathcal{Q}\openone - \qty[\mathcal{A}_1\mathcal{B}_1 + \frac{1}{\sin\delta}\qty(\mathcal{A}_1\mathcal{B}_2 + \mathcal{A}_2\mathcal{B}_1) - \frac{1}{\cos(2\delta)}\mathcal{A}_2\mathcal{B}_2]
\end{equation}
introduced in \cite{Wooltorton2022}.
\end{examplebox}

\subsection{Fixing a final optimal strategy}
The previous example box also shows how to use our method to find Tsirelson bounds saturated by strategies fixed on \emph{both} Alice's and Bob's sides.
For example, suppose you want to find Bell operators whose Tsirelson bounds are saturated by the strategy
\begin{equation}
\label{eq:max_rand_strat}
\mathcal{A}_1 = \sigma_z\,,\quad \mathcal{A}_2 = \cos (a) \sigma_z + \sin (a)\sigma_x\,, \qquad \mathcal{B}_1 = \sigma_x\,,\quad \mathcal{B}_2 = \cos (b) \sigma_z - \sin (b)\sigma_x\,.
\end{equation}
This choice is relevant because the correlation between the measurements $\mathcal{A}_1$ and $\mathcal{B}_1$ vanishes, yielding $\ev*{\mathcal{A}_1 \mathcal{B}_1}{\phi^+}=0$. 
Consequently, certifying this strategy guarantees that the measurement outcomes can be used to generate maximal global randomness in the scenario with two inputs and two outputs.
A particular case is the choice $b=\delta$ and $a=\delta-\pi/2$, which corresponds to the measurements saturating \eqref{eq:woolt_sos} and are self-tested in case where $\ev{I_\delta}=\mathcal{I}_{\mathcal{Q}}'$.

To check the possibility of deriving a boundary saturated by the previous measurements, we solve the system of equations
\begin{equation}
(K{\boldsymbol{\mathcal A}})_j =  (O \Gamma {\boldsymbol{\mathcal{B}}})_j^T \quad \forall j\,,
\end{equation}
in terms of $K,O$ and $\Gamma$. 
A solution is found by setting
\begin{equation} 
\begin{aligned}
     \gamma &= \operatorname{arccot}\qty(-\frac{\cos(a)\sin(a+b)}{\sin(b)})\,, \\
    \theta &= \arctan(-1-\cot(b)\tan(a))\,,
\end{aligned}
\qquad 
\begin{aligned}
    |\vec u_1^B| &= \tan(b)\sqrt{-\cot(b)\tan(a+b)}\,,\\
     |\vec u_2^B| &= \sqrt{\frac{\cos(a)\cos(a+b)}{\cos(b)}}\,,
\end{aligned} 
\end{equation}
provided that $-\pi/2 < a+b<0$.
The parameters defined above characterize a family of Tsirelson bounds saturated by the strategy in \eqref{eq:max_rand_strat}, and may therefore be utilized in device-independent protocols for global randomness generation; a formal study of their certification properties remains an objective for future works.
 
\subsection{Asymmetric scenarios and generalization of the elegant Bell inequality}
Here we leverage the discussion of Sec.\ \ref{sec:scenarios} to retrieve the Bell operator of the Elegant Bell Inequality (EBI) \cite{Gisin:2007gps}, defined in an asymmetric scenario, as a particular case of a more general and much broader family of operators with a symmetric number of inputs for Alice and Bob.

We consider the case in which Alice's measurements consist of four qubit observables forming the vertices of a regular tetrahedron on the Bloch sphere
\begin{equation}
\label{eq:tetrahedron}
\begin{aligned}
\iOP{\mathcal{A}}_1 &= \frac{\sigma_x - \sigma_y + \sigma_z}{\sqrt{3}}\equiv \vec n_1^A \cdot \vec \sigma\\
\iOP{\mathcal{A}}_2 &= \frac{\sigma_x + \sigma_y - \sigma_z}{\sqrt{3}}\equiv \vec n_2^A \cdot \vec \sigma
\end{aligned}\qquad
\begin{aligned}
\iOP{\mathcal{A}}_3 &= \frac{-\sigma_x - \sigma_y - \sigma_z}{\sqrt{3}} \equiv \vec n_3^A \cdot \vec \sigma\\
\iOP{\mathcal{A}}_4 &= \frac{-\sigma_x + \sigma_y + \sigma_z}{\sqrt{3}} \equiv \vec n_4^A \cdot \vec \sigma
\end{aligned}
\end{equation}
satisfying the linear dependence: $\sum_{k=1}^4 \iOP{\mathcal{A}}_k = 0$.
These measurements are known to be included in the optimal strategy of the EBI \cite{Andersson2017EBI}.

We take the above measurements as initial ones for our procedure, we define $\iOP{\mathcal{B}}_k \equiv (\iOP{\mathcal{A}}_k)^T$ and we consider the family of Bell operators defined by four dimensional orthogonal transformations, $O$.
As discussed in Sec.\ \ref{sec:qubit_strate}, orthogonal transformations are always valid, also in the case in which the initial measurements are linearly dependent. 
When we consider the $K$ contribution and we apply $O^{-1}$, we get the new operators
\begin{equation}
    \fOP{B}_j = \sum_{\ell=1}^4 (O^{-1})_{j\ell}\, \sqrt{2} k_\ell\iOP{\mathcal{B}}_\ell =  \qty[\sum_{\ell=1}^4 (O^{-1})_{j\ell}\, \sqrt{2} k_\ell\,\vec n^B_\ell] \cdot \vec\sigma \equiv \vec u_j \cdot \vec \sigma\,
\end{equation}
with $\vec u_j$ a real not normalized vector.
The Bell operator we find is then written as
\begin{equation}
    S = \openone - \sqrt{2} \sum_{\mu,\ell=1}^{4} k_\mu O_{\mu,\ell}|\vec u_\ell|\,\mathcal{A}_\mu \mathcal{B}_\ell\,.
\end{equation}
This Bell operator is defined in a symmetric scenario where Alice and Bob both have four dichotomic measurements with two possible outcomes $\pm1$.
The Bell operator is saturated $\langle S\rangle=0$ when Alice chooses the measurements given in \eqref{eq:tetrahedron}, while Bob chooses the measurement $\iOP{\mathcal{B'}}_\ell=\frac{1}{|\vec u_\ell|}\vec u_\ell\cdot\vec\sigma$.

To modify the above operator for an asymmetric scenario in the number of inputs between Alice and Bob, we need to impose the commutator
\begin{equation}
    \qty[\fOP{B}_j, (\fOP{B}_k)^\dagger]=\qty[\vec u_j \cdot \vec \sigma, \vec u_k \cdot \vec \sigma] = 2i(\vec u_j \times \vec{u}_k)\cdot \vec\sigma =0
\end{equation}
for some choice of $j,k$.
As discussed in Sec.\ \ref{sec:scenarios}, the number of measurements is determined by the number of sets of commuting operators in the final optimal strategy.
From this we conclude that $O$ is characterized by the equation $\vec u_j \times \vec{u}_k=0$, which implies that the two vectors are collinear or that at least one of them is the zero vector.
A simple choice reducing Bob's number of measurements is to satisfy directly the linear dependence described below \eqref{eq:tetrahedron} and choose $O^{-1}$ with one row proportional to the vector \( (1, 1, 1, 1) \). 
This would produce a null operator in the final optimal strategy.
More general solutions also exist, and one such example gives precisely the EBI, presented in the box at the end of this section.

We finally note that we can at most find a pair of operators to commute. 
Indeed, the transformation to the new vectors $\vec u_{j}$ is governed by a $4 \times 4$ orthogonal matrix. 
Because $O$ is full-rank (it is invertible), it preserves the dimension of the space. 
Therefore, the four new vectors $\{\vec{u}_{1}, \vec{u}_{2}, \vec{u}_{3}, \vec{u}_{4}\}$ must also span a 3-dimensional space, as the original $\vec{a}_k$.
If we had two separate pairs of commuting operators (e.g., $\vec{u}_1 \propto \vec{u}_2$ and $\vec{u}_3 \propto \vec{u}_4$), all four vectors would be constrained to just two lines, which contradicts the requirement that the vectors must span a 3-dimensional space.

\begin{examplebox}{Elegant Bell-inequality}
In the Bell scenario of the EBI \cite{Andersson2017EBI}, Alice has four possible dichotomic measurements, while Bob has three possible dichotomic measurements (we swap the role of Alice and Bob with respect to \cite{Andersson2017EBI}).
The Bell operator of the EBI satisfies:
\begin{equation}
\begin{split}
    \openone - \frac{1}{4\sqrt{3}}\Bigl[\mathcal{A}_1(\mathcal{B}_1&+\mathcal{B}_2 +\mathcal{B}_3) + \mathcal{A}_2(\mathcal{B}_1-\mathcal{B}_2-\mathcal{B}_3) + \Bigr.\\
    \Bigl.&+\mathcal{A}_3 (-\mathcal{B}_1+\mathcal{B}_2-\mathcal{B}_3)+ \mathcal{A}_4(-\mathcal{B}_1-\mathcal{B}_2+\mathcal{B}_3)\Bigr] \succeq 0\,.
\end{split}
\end{equation}
To reproduce it, we start from the optimal Alice's measurements defined by the observables \eqref{eq:tetrahedron}.
We then consider 
\begin{equation}
    K = \frac{1}{2} \openone_4\,, \qquad O^{T} = \frac{1}{2}\begin{pmatrix}
1 & 1 & -1 & -1 \\
1 & -1 & 1 & -1 \\
\sqrt{2} & 0 & 0 & \sqrt{2} \\
0 & \sqrt{2} & \sqrt{2} & 0
\end{pmatrix}\,,
\end{equation}
from which we obtain the new observables
\begin{equation}
\begin{aligned}
\fOP{B}_{1} &= \frac{1}{\sqrt{3}}\sigma_x \equiv \frac{1}{\sqrt{3}}\fOP{\mathcal B}_{1} \ , \\
\fOP{B}_{2} &= \frac{1}{\sqrt{3}}\sigma_y \equiv \frac{1}{\sqrt{3}}\fOP{\mathcal B}_{2}
\end{aligned}
\qquad
\begin{aligned}
\fOP{B}_{3} &= \frac{1}{\sqrt{6}}\sigma_z \equiv \frac{1}{\sqrt{6}}\fOP{\mathcal B}_{3}\\ 
\fOP{B}_{4} &= -\frac{1}{\sqrt{6}}\sigma_z  \equiv \frac{1}{\sqrt{6}}\fOP{\mathcal B}_{4} =- \fOP{B}_{3} \,.
\end{aligned}
\end{equation}
From the above equations we identify $\Gamma = \text{diag}\qty(1/\sqrt{3},1/\sqrt{3},1/\sqrt{6},1/\sqrt{6})$, and with $K$ and $O$ we obtain a Bell operator, written in terms of unitary operators, and defined in a symmetric scenario with $m=4$ dichotomic measurements:
\begin{equation}
\begin{split}
    \openone-{\boldsymbol{\mathcal A}}^\dag KO\Gamma {\boldsymbol{\mathcal B}} = \openone - \frac{1}{4\sqrt{3}}\Bigl[&\mathcal{A}_1(\mathcal{B}_1+\mathcal{B}_2 +\mathcal{B}_3) + \mathcal{A}_2(\mathcal{B}_1-\mathcal{B}_2+\mathcal{B}_4) + \Bigr.\\
    \Bigl.&+\mathcal{A}_3(-\mathcal{B}_1+\mathcal{B}_2+\mathcal{B}_4) + \mathcal{A}_4(-\mathcal{B}_1-\mathcal{B}_2+\mathcal{B}_3)\Bigr]\succeq 0\,,
\end{split}
\end{equation}
saturated by the strategy
\begin{equation}
\begin{aligned}
\fOP{\mathcal{A}}_1 &= \frac{\sigma_x - \sigma_y + \sigma_z}{\sqrt{3}}\\
\fOP{\mathcal{A}}_2 &= \frac{\sigma_x + \sigma_y - \sigma_z}{\sqrt{3}}
\end{aligned}\quad
\begin{aligned}
\fOP{\mathcal{A}}_3 &= \frac{-\sigma_x - \sigma_y - \sigma_z}{\sqrt{3}} \\
\fOP{\mathcal{A}}_4 &= \frac{-\sigma_x + \sigma_y + \sigma_z}{\sqrt{3}}
\end{aligned}\qquad 
\begin{aligned}
\fOP{\mathcal B}_{1} &= \sigma_x\\
\fOP{\mathcal B}_{2} &= \sigma_y
\end{aligned}
\quad
\begin{aligned}
\fOP{\mathcal B}_{3} &= \sigma_z\\ 
\fOP{\mathcal B}_{4} &= -\sigma_z
\end{aligned} \ . 
\end{equation}
However, the existence of an optimal strategy in which the projectors of two measurements are the same ($\qty[\fOP{\mathcal{B}}_{4},\fOP{\mathcal{B}}_{3}]=0$) motivates the reduction of the number of measurements in the final Bell scenario.
In this case, the equality $\fOP{\mathcal{B}}_{4} =- \fOP{\mathcal{B}}_{3}$ allows for the redefinition of $\mathcal{B}_4 = -\mathcal{B}_3$ also in terms of the symbolic variables expressing the final Bell operator, which reduces to the one of the EBI.
\end{examplebox}

\section{Higher-dimensional inequalities}
\label{sec:high_dim}
As the dimension $d$ of the Hilbert space for the optimal strategy increases, the landscape of available quantum correlations becomes significantly more complex, as do the potential relations between operators of the same or different inputs. Hence, complete and fully general arguments become difficult to carry out.
Nevertheless, in this section, we discuss solutions that are valid for any number of Alice's inputs and any choice of her strategy. We then restrict our focus to the case of two inputs ($m=2$), while also introducing constraints on Bob's side.

\subsection{General solutions}
Let us first recap what we found in the qubit case.
In $d=2$, any traceless normal operator is proportional to an Hermitian traceless operator. 
By using the unitary invariance, we can eliminate irrelevant phases and choose Alice's strategy as made exclusively with Hermitian operators.
When we want to construct nontrivial Bell operators, we combine the transpose of these initial Hermitian operators (which are still Hermitian). 
Now, a generic linear combination of independent Hermitian operators 
remains Hermitian if and only if the coefficients are real. 
In the bidimensional case, this requirement restricted the transformations to real coefficients (up to potential linear dependencies) when combining operators from different inputs in the initial optimal strategy to identify a new input in the final optimal strategy. 
For this reason, to obtain relevant Bell operators, we restricted the transformations from the unitary group to the real orthogonal group.

In $d$ dimensions, the structure is more complex. Each traceless normal operator can be expressed as a complex combination of $(d-1)$ commuting, traceless Hermitian operators. 
This can be seen as a direct consequence of the spectral decomposition and the completeness relation of the set of projectors, and allows in principle to a much richer structure than the $d=2$ case. Simple arguments like the one around equation \eqref{eq:qubit_norm_cond} do not apply, and in principle also complex transformations play a role.

However, while the $d$-dimensional structure allows for more intricate algebraic relations, it remains true that a real linear combination of Hermitian operators is itself Hermitian (and therefore diagonalizable), representing a well-defined measurement input.
Consequently, we can generalize the qubit construction and find a  solution in $d$ dimensions considering
real matrices
\begin{equation}
(DW)_{ij} \in \mathds{R} \quad \text{and} \quad U\equiv O \in \text{O}(md)\,,
\end{equation}
to obtain the valid sum of squares decomposition:
\begin{equation}
\label{eq:qudit_gen}
 S = \mathbf{G}^\dag \mathbf{G}\,,
    \qquad \mathbf{G}=DW{\boldsymbol{\Lambda}}-O\Gamma{\boldsymbol{\Pi}}\,,
\end{equation}
whose Tsirelson bound is saturated by a generic choice of Alice's strategy $\iOP{\mathbf{\Lambda}}$ and a corresponding strategy for Bob $\iOP{\mathbf{\Pi}}$
defined by the projectors of the operators $ \iOP{\bf{B}}=O^{-1} (\mathbf{\iOP{\Lambda}}^T) $ (where the transpose is on the operators acting on the Hilbert space of the state).
This formula is the high-dimensional analogue of the recipe we gave for the qubit case in Sec.~\ref{sec:qubit_strate}. In that case, we also used the Bloch vector representation to simplify our results: the generalization to $d$ dimensions requires substituting the Pauli matrices with the $d^2-1$ generalized Gell-Mann matrices. 
It is important to note, however, that unlike the qubit case, not every vector in $\mathbb{R}^{d^2-1}$ corresponds to an operator with a valid physical spectrum, as the geometry of the state space becomes more complex in higher dimensions \cite{Bertlmann:2008jgt}.

\paragraph{More scenarios}
The number of inputs in Bob's scenario depends on the choice of matrix $DW$ and on the commutation relations of the final optimal strategy,  as we already discussed in Sec.~\ref{sec:scenarios}.
This number can be reduced if the transformed operators commute:
\begin{equation}
\qty[\fOP{B}_k, \fOP{B}_{k'}] = 0.
\end{equation}
In $d=2$ this can only happen if the two traceless operators are proportional, while in $d>2$ two traceless normal operators can commute even without being proportional. 
However, there are some constraints we need to consider.

First, the set of normal operators that define the initial strategy spans a space in the set of matrices acting on a $d$-dimensional space. 
When we apply an invertible transformation to that set, the dimension of the spanned space cannot change.\footnote{Indeed, because the transformed operators are built from the initial elements, the space spanned by the new operators is contained within the space spanned by the original operators. At the same time, since the transformation is invertible, the space spanned by the original operators is contained entirely within the space spanned by the new operators. Therefore, the two spaces must be identical.} 
This means, for example, that we cannot hope to have proportional operators in the transformed strategy unless there are some linear dependencies in the initial one.
Second, for some choices of initial strategies, it is not even possible to construct two or more \emph{non}-proportional commuting normal operators. We show this in the particular case of some mutually unbiased bases in Appendix \ref{sec:MUBs}.
The combination of these aspects restrict the possibility of obtaining particular scenarios on Bob's side, when Alice's strategy is fixed.

In the following we discuss some particular cases of the general Bell operator expressed in 
\eqref{eq:qudit_gen}. 
We start by showing, in the following box, that the inequality introduced in \cite{Tavakoli:2020zlz} is
included in this general formalism.
To the best of our knowledge, the Bell inequalities of \cite{Tavakoli:2020zlz} are the only ones in the literature which are maximally violated by two MUBs in arbitrary dimension $d>2$ \cite{McNulty2026mutuallyunbiased}. In the next section, we will propose alternatives with fewer inputs.

\begin{examplebox}{The Mutually Unbiased Bases inequalities of \cite{Tavakoli:2020zlz}}
    In \cite{Tavakoli:2020zlz}, there is an example of an inequality which is maximally saturated using two MUBs in arbitrary dimensions and admits the SOS decomposition 
    \begin{equation}
    \label{eq:SOS_Tavakoli_MUBs}
       S= \frac{1}{2}\sqrt{\frac{d-1}{d}}\sum_{x_1,x_2} \left(A^{(x_1,x_2)}\otimes\openone-\openone \otimes \sqrt{\frac{d}{d-1}}\qty(\Pi_{x_1}^{(1)}-\Pi_{x_2}^{(2)})\right)^2 \ .
    \end{equation}
    In this scenario, Alice chooses between one of $d^2$ inputs labelled by the pair $(x_1,x_2)$, each one with three outputs associated with the eigenvalues $\pm 1$ and $0$ of $A^{(x_1,x_2)}$. 
    Bob can choose between two inputs, identified with the two sets of $d$ projectors $\{\Pi^{(1)}_{x_1} \}$ and $\{\Pi^{(2)}_{x_2}\}$.
    By expanding \eqref{eq:SOS_Tavakoli_MUBs}, we obtain
    \begin{equation}
    \label{eq:Tarmin}
    I \equiv \sum_{x_1,x_2} \left( A^{(x_1,x_2)} \otimes \qty(\Pi^{(1)}_{x_1} - \Pi^{(2)}_{x_2}) \right) -\frac{1}{2}\sqrt{\frac{d-1}{d}} \sum_{x_1,x_2} \left( A^{(x_1,x_2)2} \otimes \openone \right) \preceq \sqrt{d(d-1)}\openone   
    \end{equation}
    with 
    \begin{equation}
        \mathcal{I}_{\rm LHV}=2(d-1)\left(1-\frac{1}{2} \sqrt{\frac{d-1}{d}}\right) 
    \end{equation}
    being the local bound of $I$.
    Here, we aim to reproduce the SOS decomposition \eqref{eq:SOS_Tavakoli_MUBs} as a particular case of equation \eqref{eq:qudit_gen}.
    
    To match the notation in \cite{Tavakoli:2020zlz}, we start from a scenario with $d^2$ inputs per party, labelled by an index $x$, and $d$ outcomes. 
    Looking at Alice's side in \eqref{eq:SOS_Tavakoli_MUBs}, we want a single operator per input. 
    We then force this in the initial SOS
    with a combination of only two projectors for each input
    \begin{equation}
        S_0=\sum_{x=1}^{d^2} \qty(A^{(x)}-B^{(x)})^2
        \,,\qquad
    A^{(x)}=\Lambda^{(x)}_1-\Lambda^{(x)}_2\,,\quad
    B^{(x)}=\Pi^{(x)}_1-\Pi^{(x)}_2\,.
    \end{equation}
    The above choice corresponds to consider $1\times d$ matrices $M^{(x)}$ given by $M^{(x)}=(1, -1, 0, \cdots ,0)$.
    Since we will only rotate Bob's side, we choose as Alice's initial (and final) strategy the one which is optimal in \cite{Tavakoli:2020zlz}. 
    Let's consider two sets of projectors, 
    $\{\iOP{\Pi}^{(1)}_{x_1} \}$ and $\{\iOP{\Pi}^{(2)}_{x_2}\}$ that are MUBs, namely ${\rm Tr}[\iOP{\Pi}^{(1)}_{x_1}\iOP{\Pi}^{(2)}_{x_2}]=1/d$ $\forall x_1,x_2$. 
    The optimal Alice strategy $\iOP{\Lambda}^{(x)}_{a}$ is defined by the following relation:
    \begin{equation}
        \iOP{A}^{(x)}=\sqrt{\frac{d}{d-1}}\qty(\iOP{\Pi}^{(1)}_{x_1}-\iOP{\Pi}^{(2)}_{x_2})^T 
        \equiv\iOP{\Lambda}^{(x)}_{1}-\iOP{\Lambda}^{(x)}_{2}
    \end{equation}
    where $x \equiv x_1 x_2 \in [d]^2$.
    Indeed, with two MUBs, the spectrum of the operators $\iOP{A}^{(x)}$ is $\pm 1$ with $d-2$ zero eigenvalues, as described in \cite{Tavakoli:2020zlz}.
   Defining the 
    auxiliary matrix $N_{x_1,x_2} = \frac{1}{\sqrt{d-1}}\left(\delta_{x_1,x_2} - \frac{1}{d}\right)$, the elements of the orthogonal transformation $O \in O(d^2)$ we are looking for are:
    \begin{equation}
        O_{j,x}^T = 
\begin{cases} 
\frac{1}{\sqrt{d}} \left( \delta_{j, x_1} + a + \frac{1}{\sqrt{2}} N_{x_1,x_2} \right) & \text{for } 1 \leq j \leq d \\ 
-\frac{1}{\sqrt{d}} \left( \delta_{j-d,x_2} + c + \frac{1}{\sqrt{2}} N_{x_1,x_2} \right) & \text{for } d+1 \leq j \leq 2d \\ 
Z_{j-2d,(x_1,x_2)} & \text{for } 2d+1 \leq j \leq d^2 
\end{cases}
    \end{equation}
    where $j=1,\dots, d^2$, the coefficients $a = \frac{1}{d} \left( \frac{1}{\sqrt{2}} - 1 \right)$ and $c = \frac{1}{d} \left( -\frac{1}{\sqrt{2}} - 1 \right)$ are constants required to 
satisfy orthogonality, and
 $Z_{j,x}$ are the elements of any $(d^2-2d) \times d^2$ matrix whose rows form an orthonormal basis for the subspace satisfying:
\begin{equation}
\label{eq:z_prop}
    \sum_{x_1=1}^d Z_{j, (x_1,x_2)} = 0, \quad 
\sum_{x_2=1}^d Z_{j, (x_1,x_2)} = 0, \quad 
\sum_{y=1}^d Z_{k, (y, y)} = 0 \ . 
\end{equation}
     When we apply this transformation matrix $O^T$ to the original set of operators $\iOP{B}^{(x)}$, we get three components:
    \begin{itemize}
        \item For the first block ($1 \leq j \leq d$):
        $$\sum_{x_1, x_2} \left[ \frac{1}{\sqrt{d}} \left( \delta_{j, x_1} + a + \frac{1}{\sqrt{2}} N_{x_1, x_2} \right) \right] \sqrt{\frac{d}{d-1}} (\iOP{\Pi}^{(1)}_{x_1} - \iOP{\Pi}^{(2)}_{x_2})=\frac{1}{\sqrt{d-1}} (d \iOP{\Pi}^{(1)}_{j}-\openone) \ .$$
    \item For the second block ($d+1 \leq j \leq 2d$):
    $$ \sum_{x_1, x_2} \left[ -\frac{1}{\sqrt{d}} \left( \delta_{j-d, x_2} + c + \frac{1}{\sqrt{2}} N_{x_1, x_2} \right) \right] \sqrt{\frac{d}{d-1}} (\iOP{\Pi}^{(1)}_{x_1} - \iOP{\Pi}^{(2)}_{x_2})=\frac{1}{\sqrt{d-1}} (d \iOP{\Pi}^{(2)}_{j-d}-\openone) \ . $$
    \item For the third block ($2d+1 \leq j \leq d^2$):
    $$\sum_{x_1, x_2} Z_{j-2d, (x_1, x_2)} \sqrt{\frac{d}{d-1}} \qty(\iOP{\Pi}^{(1)}_{x_1} - \iOP{\Pi}^{(2)}_{x_2})=0$$
    where we used the properties \eqref{eq:z_prop}.
    \end{itemize}
    These expressions immediately tell us that Bob's strategy is made of two sets of projectors, $\iOP{\Pi}^{(1)}$ and $\iOP{\Pi}^{(2)}$ while the other possible inputs collapse to zero.
    When $S_0$ is transformed with $O$, we get our targeted SOS decomposition \eqref{eq:SOS_Tavakoli_MUBs}.
\end{examplebox}

\subsection{Two inputs on Alice's side}
\label{sec:qudit_orth}
This section provides an extended treatment of the scenario with two inputs on Alice's side. 
We first consider a simple application of formula \eqref{eq:qudit_gen}, leading to a scenario with $2d$ inputs on Bob's side. 
We discuss in particular detail the case in which Alice's strategy is constructed from MUBs providing a Bell operator with fewer inputs than the one of \cite{Tavakoli:2020zlz}. 
Then, we discuss more general solutions than \eqref{eq:qudit_gen} that reduce the number of Bob's inputs, allowing us to reproduce and generalize the SATWAP operator for two inputs.

\subsubsection{$m_A=2$ and $m_B=2d$}
We consider formula \eqref{eq:qudit_gen} choosing $DW=\openone$, so that Alice's optimal strategy simply consists of two sets ($m_A=2$) of $d$ orthogonal rank-1 projectors, with $d >2$.
We apply a simple $2d$-dimensional rotation:\begin{equation}
O = 
\begin{pmatrix}\cos\theta \openone_d & -\sin\theta  \openone_d \\
\sin\theta \openone_d & \cos\theta  \openone_d\end{pmatrix}
\end{equation}
and obtain the SOS decomposition
\begin{equation}
\label{eq:S_rot_d}
S =  \sum_{k=1}^{d} \left[ \left( \Lambda_k^{(1)} - \cos\theta B_k + \sin\theta B_{k+d} \right)^2 + \left( \Lambda_k^{(2)} - \sin\theta B_k - \cos\theta B_{k+d} \right)^2 \right]\end{equation}
where each $B_k$ or $B_{k+d}$, with $k=1, \dots, d$, is an operator associated with an input.
Expanding this SOS we find:
\begin{equation}
    2\openone-\sum_{k=1}^{d} \Bigg[-B_k^2 - B_{k+d}^2 + 2\left(\cos\theta \Lambda_k^{(1)} + \sin\theta \Lambda_k^{(2)}\right)B_k + 2\left(-\sin\theta \Lambda_k^{(1)} + \cos\theta \Lambda_k^{(2)}\right)B_{k+d} \Bigg] \succeq 0\ .
\end{equation}
The Bell inequality is completely defined when the eigenvalues of the operators $B_k$ are specified. As discussed in the general method,
every choice of Alice's optimal projectors $\bf{\iOP{\Lambda}}$ corresponds to a determined 
choice of the $B_k$'s eigenvalues, as detailed below.
Denoting Bob's initial strategy by $\iOP{\bf{\Pi}}$, which is the transpose of Alice's, his final optimal operators are given by:
\begin{equation}
\label{eq:transf_op_qudit}
\begin{aligned}
\fOP{B}_k &=
\cos\theta \, \iOP{\Pi}^{(1)}_k + \sin\theta \, \iOP{\Pi}^{(2)}_k 
\\
\fOP{B}_{k+d} &=-\sin\theta \, \iOP{\Pi}^{(1)}_{k} + \cos\theta \, \iOP{\Pi}^{(2)}_{k}
\end{aligned}\,,
\quad\qquad k=1,\cdots, d \ .
\end{equation}
Because these operators are constructed from two rank-1 projectors, their total rank can never exceed 2, regardless of the overall dimension $d$. 
This means that each normal operator has only $2$ non-vanishing eigenvalues. 
By defining
\begin{equation}
    \Tr(\iOP{\Pi}^{(1)}_j \iOP{\Pi}^{(2)}_j) =  \Tr(\iOP{\Lambda}^{(1)}_j \iOP{\Lambda}^{(2)}_j) = \abs{c_{j}}^2 
\end{equation}
the eigenvalues of the operators $\fOP{B}_k$
and $\fOP{B}_{k+d}$ are found to be
\begin{equation}
\label{eq:eigen_qudit}
\begin{aligned}
    \lambda^{(k)}_{\pm}&=\frac{1}{2} \left( \cos\theta + \sin\theta \pm \sqrt{1 - \sin(2\theta)\qty(1 - 2\abs{c_k}^2)} \right)  \\
    \lambda^{(k+d)}_{\pm}&=\frac{1}{2} \left( \cos\theta - \sin\theta \pm \sqrt{1 + \sin(2\theta)\qty(1 - 2\abs{c_{k}}^2)} \right)
\end{aligned}
\end{equation}
where $\lambda^{(k)}_{\pm}$ ($\lambda^{(k+d)}_{\pm}$) are the two non-vanishing eigenvalues of $\fOP{B}_k$
($\fOP{B}_{k+d}$) and $k=1,\cdots,d$.
If we now express every operator $B_k$ in its 
spectral decomposition $B_k=\sum_b\lambda^{(k)}_b\Pi^{(k)}_b$, the above 
Bell inequality can be written in terms of 
the symbolic projectors $\{\Lambda^{(x)}_a,\Pi^{(k)}_b\}$:
\begin{equation}
\label{eq:qudit_expanded}
\begin{aligned}    
S=2\openone-2&\sum_{k=1}^{d}\sum_{b=\pm} 
\Big[\lambda^{(k)}_b
\left(\cos\theta \Lambda_k^{(1)} + \sin\theta \Lambda_k^{(2)}-\frac12\lambda^{(k)}_b
\right)\Pi^{(k)}_b +
\\
&
+\lambda^{(k+d)}_b
\left(-\sin\theta \Lambda_k^{(1)} + \cos\theta \Lambda_k^{(2)}-\frac12\lambda^{(k+d)}_b
\right)\Pi^{(k+d)}_b
\Big] \succeq 0\ .
\end{aligned}
\end{equation}

\paragraph{The MUB case} 
A simple application of \eqref{eq:qudit_expanded} is obtained by choosing two mutually unbiased bases on Alice's side. 
The defining geometric property of MUBs in $d$ dimensions is that the overlap probability scales inversely with the dimension, so $\abs{c_k}^2= 1/d$. 
With this substitution, the eigenvalues of the two blocks in \eqref{eq:eigen_qudit} become independent of the input $k$:
\begin{align}
    \lambda_{\pm}^{(k)}(\theta)&=
     \frac{1}{2} \left( \cos\theta + \sin\theta \pm \sqrt{1 - \sin(2\theta)\qty(1 - 2/d)} \right) \ , \qquad \\
     \lambda_{\pm}^{(k+d)}(\theta)&=
    \frac{1}{2} \left( \cos\theta - \sin\theta \pm \sqrt{1 + \sin(2\theta)\qty(1 - 2/d)} \right)\,.
\end{align}

Note that our construction yields a Bell operator with two inputs and $d$ outcomes on Alice's side, and $2d$ inputs and $d$ outcomes on Bob's side (of which $d-2$ are null outcomes). 
This result can be compared with the one in \cite{Tavakoli:2020zlz} (reproduced in equation \eqref{eq:SOS_Tavakoli_MUBs}), which instead requires $d^2$ inputs on one of the two sides.
The (im)possibility of constructing Bell operators with even fewer inputs on Bob's side for generic MUBs is discussed in Appendix \ref{sec:MUBs}.

For a better comparison  with the results in the literature, we also need to understand the relation between quantum and classical bounds of \eqref{eq:qudit_expanded}. 
For simplicity we only discuss the case $\theta=\pi/4$.
Define
\begin{equation}
\label{eq:T}
    I=2\openone-S \ .
\end{equation}
From the SOS decomposition, we know that the quantum bound of $I$ is $\mathcal{I}_\mathcal{Q} = 2$ while the local bound is given by (see Appendix \ref{app:d_local_bound})
\begin{equation}
    \mathcal{I}_{\rm LHV} = \max \left[ \frac{3}{2} + \frac{1}{\sqrt{d}} - \frac{1}{2d}, 2\sqrt{1 - \frac{1}{d}} \right]\,,
\end{equation}
where the first term is larger when $d\le 5$ and the second is larger when $d \ge 6$.
Since both terms are smaller than $\mathcal{I}_\mathcal{Q}=2$, this result shows the possibility, in principle, to use the Bell operator \eqref{eq:S_rot_d} for proper quantum protocols.

\begin{figure}
    \centering
    \includegraphics[width=0.6\linewidth]{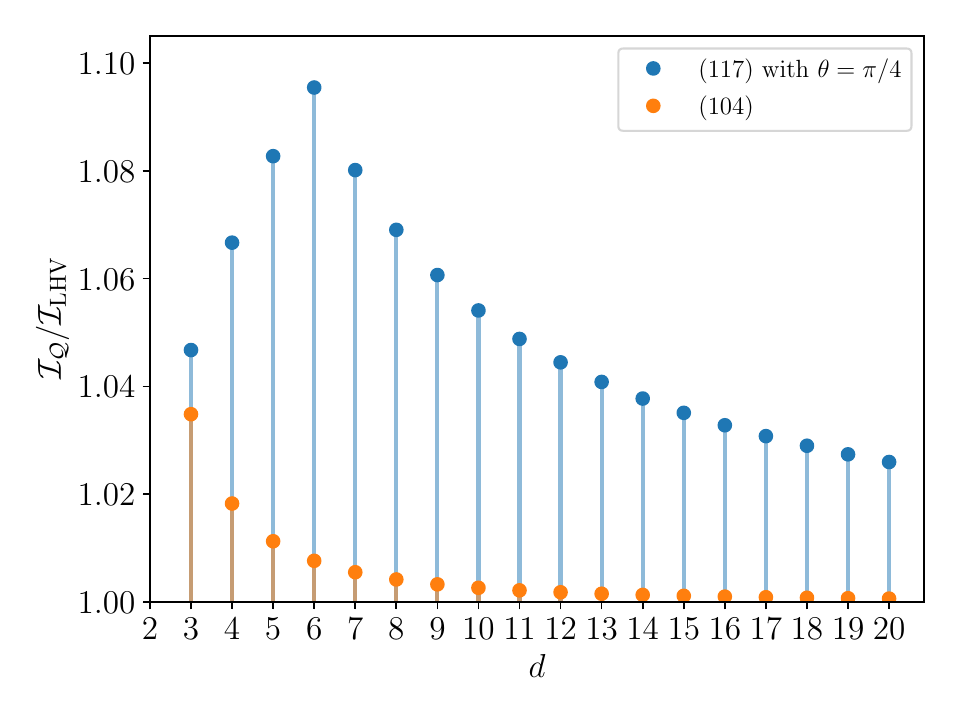}
    \caption{Comparison of the quantum-to-local bound ratios for the Bell operator \eqref{eq:Tarmin} from \cite{Tavakoli:2020zlz} and the novel Bell operator \eqref{eq:T} derived in this work.}
    \label{fig:comparison}
\end{figure}

Figure~\ref{fig:comparison} displays the quantum-to-local bound ratio,
$\mathcal{I}_{\mathcal Q}/\mathcal{I}_{\rm LHV}$, for the Bell operators
defined in Eqs.~\eqref{eq:Tarmin} (from Ref.~\cite{Tavakoli:2020zlz})
and~\eqref{eq:T}. We find that, for all $d \ge 3$, the operator in
Eq.~\eqref{eq:T} achieves a strictly larger ratio.

\subsubsection{The $m_A=m_B=2$ symmetric scenario}
\label{sec:d_m2}

We now try to find Bell operators in which Bob’s strategy also involves two inputs.
This investigation is driven by two primary motivations.
First, this specific configuration has been previously explored in the literature (e.g., \cite{SATWAP2017, Sarkar2021}). We aim to verify if these results can be reproduced and extended with our techniques (a question we answer in the affirmative).
Second, this scenario yields transformation examples that go beyond the orthogonal ones previously introduced and it allows us to examine the requirement of commutativity between sets of transformed variables. Beyond the specific findings, this serves to show some techniques that can be used to derive explicit transformation matrices when specific constraints are placed on Bob’s scenario.

Since this section is technically dense, its structure is summarized here to facilitate navigation through the results. 
The analysis begins by identifying the specific measurements on Alice's side required to obtain the symmetric scenario, leading to Equation \eqref{eq:cauchy_form} and to the choice of the CGLMP measurements in Eq.\ \eqref{eq:Alice_CGLMP}. After reformulating the problem into a more convenient framework in Eqs.\ \eqref{eq:cons_1} and \eqref{eq:cons_2}, the generalized version of the SATWAP operator for two inputs is derived in Eq.\ \eqref{eq:gen_SATWAP_op}.

\emph{Comment on the notation:} In this last section we will not need to use different subscripts for normal operators and for projectors.
We will instead use letters $i,j,k, \dots$ in the range from $0$ to $d-1$ (instead of $1$ to $d$) to match the existing literature and simplifying the notation.

\paragraph{Choosing Alice's measurements}
As we learn from the previous discussion and Appendix \ref{sec:MUBs}, a generic strategy on Alice's side will in general lead to a Bell operator with $2d$ inputs on Bob's side without the possibility of reducing this number to only two inputs. 
So, we need to look for some guidance to understand which specific strategy to choose for Alice to achieve this scenario.

Remember that the operators in Bob's optimal strategy are combinations of the transpose of the $2d$ projectors, $\iOP{\Lambda}_j^{(x)}=\ketbra*{\alpha^{(x)}_j}$, which define Alice's strategy.
To have only two inputs these operators must form two sets of $d$ commuting operators and a natural idea is to require, for each set, these operators to be independent, which is typical for self-testing strategies. 
This means that they can be combined to produce a rank-one projector $\fOP{\Pi}=\ketbra*{v}$ that can be written as
\begin{equation}
    \fOP{\Pi}=\ketbra*{v}=\sum_j b_j^{(1)} \ketbra*{\alpha_j^{(1)}}+\sum_j b_j^{(2)} \ketbra*{\alpha_j^{(2)}} 
\end{equation}
for some coefficients $b_j^{(x)}$. 
Now, we project this operator equation into the mixed basis by sandwiching it between $\bra*{\alpha_k^{(1)}}$ and $\ket*{\alpha_{\ell}^{(2)}}$:
\begin{equation}
    \braket*{\alpha_k^{(1)}}{v}\braket*{v }{\alpha_\ell^{(2)}} = b_k^{(1)} \braket*{\alpha_k^{(1)}}{\alpha_\ell^{(2)}} + b_\ell^{(2)} \braket*{\alpha_k^{(1)}}{\alpha_\ell^{(2)}} \ .
\end{equation}
Factoring out the overlap matrix element $c_{k\ell}^{(12)} \equiv \braket*{\alpha_k^{(1)}}{\alpha_\ell^{(2)}}$, we find:
\begin{equation}
\label{eq:cauchy_form}
    c_{k\ell}^{(12)} = \frac{\braket*{\alpha_k^{(1)}}{v}\braket*{v}{\alpha_\ell^{(2)}}}{b_k^{(1)} + b_\ell^{(2)}} \ . 
\end{equation}
The expression we just found has the structure of a Cauchy-like matrix: the dependence on the rows and columns factorizes in the numerator, while in the denominator it appears through a sum.
So, to obtain the scenario we have in mind, with two inputs on both parties, and $d$ independent operators for each input of Bob's optimal strategy, it is necessary that the two bases associated with Alice's optimal strategy are related by a Cauchy-like transformation.
Moreover, $c_{ij}$ must also be unitary, being a change of basis between two orthonormal bases. 
A remarkable example satisfying these requirements is given by the measurements used in the CGLMP inequalities \cite{Collins:2001qdi}:
\begin{equation}
\label{eq:Alice_CGLMP}
    {\iOP{A}_{j}^{(x)}}=\sum_{k=0}^{d-1}\omega^{kj} \ketbra*{\alpha^{(x)}_k} \ , \qquad  \ket*{\alpha^{(x)}_k}=\frac{1}{\sqrt{d}}\sum_{q=0}^{d-1}\omega^{(k-\frac{x}{2}+\frac{1}{4})q}\ket*{q}
\end{equation}
where $\omega=e^{2\pi i/d}$ and $x=1,2$, 
which is characterized by the change of basis matrix
\begin{equation}
    c_{jk}^{(12)} =\frac{1}{d} \sum_{q=0}^{d-1} \omega^{-(j-k + 1/2)q}=\frac{2}{d}\left(\frac{1}{1-\omega^{k-j-1/2}}\right) 
\end{equation}
with the manifest Cauchy-like structure.
Considering this as Alice's strategy, we now discuss which class of Bell operators can be found.  

\paragraph{Rephrasing the commutation problem}
As said, we want two sets of $d$ commuting operators and, up to an irrelevant permutation, we can always assume that the commutation happens between the first $d$ and the second $d$ operators.
This structure mirrors the commutation property of Bob's initial strategy, $\iOP{B}_j^{(x)}=(\iOP{A}_j^{(x)})^T$, and motivates a small change in our notation.
We will divide the unitary matrix $U$ into four blocks and equip it with two kinds of indices so that we will write it $(U)_{ij}^{(xy)}$.
Indices on the top $x,y=1, \dots m$ are related to the inputs and indices on the bottom $i,j=0,\dots,d-1$ are associated with different outputs of the same input.
With this notation the transformed operators in \eqref{eq:AVA} are
\begin{equation}
    \fOP{B}^{(y)}_k=\sum_{y'}\sum_{k'}(U)_{kk'}^{(yy')}{\iOP{B}}_{k'}^{(y')}
\end{equation}
and the commutation constraints are
\begin{equation}
\label{eq:comm_eq_for_SATWVAP}
     \qty[\fOP{B}_k^{(y)},\qty(\fOP{B}_{k'}^{(y)})^\dagger]=     \sum_{i,j=0}^{d-1} \sum_{x, z=1}^2 U_{ki}^{(yx)}\qty(U_{k'j}^{(y z)})^* \qty[\iOP{B}_i^{(x)},\qty(\iOP{B}_j^{(z)})^\dagger] = 0 \ , 
\end{equation}
where the $\iOP{B}^{(x)}_j$ are the transposes of the matrices $\iOP{A}_j^{(x)}$ in \eqref{eq:Alice_CGLMP}.
Note that the matrix $M$, defining the operators $\iOP{B}_j^{(x)}$ in terms of projectors, is (up to an irrelevant rescaling) unitary: 
\begin{equation}
    M^{(x)}_{jk}=\omega^{jk} \ .
\end{equation}
Hence it can be just interpreted as a change of basis from the projectors to the normal operators. 
Since we can formulate the problem in the more convenient basis, we choose the projector basis and perform the change of basis at the end.
So, in equation \eqref{eq:comm_eq_for_SATWVAP}, we will replace the normal operators $\iOP{B}^{(x)}_i$ and $\iOP{B}^{(z)}_j$ with the corresponding projectors.

The central objects to analyze are the commutators between the projectors of the two bases of the initial strategy. Setting $\iOP{\Pi}_i^{(x)}=\ketbra*{\beta_i^{(x)}}$:  
\begin{equation}
    \qty[\iOP{\Pi}_i^{(x)},\iOP{\Pi}_j^{(z)}]=c_{ji}^{(zx)}\ketbra*{\beta_i^{(x)}}{\beta_j^{(z)}}-c_{ij}^{(xz)}\ketbra*{\beta_j^{(z)}}{\beta_i^{(x)}} \ ,
\end{equation}
where we note that, since $\iOP{\Pi}_i^{(x)}=(\iOP{\Lambda}_i^{(x)})^T$, then $\braket*{\beta_i^{(x)}}{\beta_j^{(z)}}=\braket*{\alpha_j^{(z)}}{\alpha_i^{(x)}}={c_{ji}^{(zx)}}$. 
Considering the bracket with $\bra*{\beta_{\alpha}^{(x)}}$ and $\ket*{\beta_{\beta}^{(x)}}$ ($\alpha,\beta=0,\dots,d-1$), gives the components
\begin{equation}
\label{eq:comm_comp}
  \bra*{\beta_{\alpha}^{(x)}}\qty[\iOP{\Pi}_i^{(x)},\iOP{\Pi}_j^{(z)}]\ket*{\beta_{\beta}^{(x)}} = c_{ji}^{(zx)}c_{\beta j}^{(xz)} \delta_{\alpha i} - c_{ij}^{(xz)} c_{j \alpha}^{(zx)} \delta_{i \beta}\equiv (C_{\alpha \beta})_{ij}^{(xz)} \ .
\end{equation}
So we can reformulate the problem as
\begin{equation}
\label{eq:constr_commutator_m2}
    \sum_{i,j=1}^d \sum_{x, z=1}^2 U_{ki}^{(yx)}\qty(U_{k'j}^{(y z)})^* \qty(C_{\alpha \beta})_{ij}^{(xz)} = 0 \ .
\end{equation}
In matrix form, this choice allows us to write the constraint \eqref{eq:constr_commutator_m2} as the requirement that the transformed matrices have vanishing block-diagonal components:
\begin{equation}
\label{eq:comm_const_gen}
\qty(
U \, (C_{\alpha \beta}) \, U^\dagger
)^{(yy)}_{kk'}
=
0,
\qquad
\forall\, y,k,k',\alpha,\beta.
\end{equation}
Using the algebra of $U(dm)$, the transformation can also be expressed in terms of a Hermitian matrix $X$ such that $U=e^{i X}$:
\begin{equation}
\label{eq:const_exp}
\qty(e^{iX} \, (C_{\alpha\beta}) \, e^{-iX})^{(yy)}_{kk'}
=0 \qquad
\forall\, y,k,k',\alpha,\beta.
\end{equation}
We decompose the matrix $X$ into a block-diagonal and a block anti-diagonal component,
\begin{equation}
    X=\begin{pmatrix}
        X^{(11)} & X^{(12)} \\
        X^{(21)} & X^{(22)}
    \end{pmatrix}
    =
    \begin{pmatrix}
        X^{(11)} & 0 \\
        0 & X^{(22)}
    \end{pmatrix}
    +
    \begin{pmatrix}
        0 & X^{(12)} \\
        X^{(21)} & 0
    \end{pmatrix}
    \equiv X_1+X_2 .
\end{equation}
Here $X^{(11)}$ and $X^{(22)}$ are Hermitian matrices, while $X^{(12)}=(X^{(21)})^\dagger$ is arbitrary. Each block $X^{(xy)}$ carries indices $i,j$.
To simplify the problem, we make the ansatz that the block-diagonal and anti-block-diagonal parts commute,
\begin{equation}
\label{eq:ansatz_comm}
    [X_1,X_2]=0 \ .
\end{equation}
Under this assumption,
\begin{equation}
    e^{iX} \, (C_{\alpha \beta}) \, e^{-iX}
    =
    e^{iX_1}e^{iX_2} \,(C_{\alpha \beta}) \, e^{-iX_2}e^{-iX_1}.
\end{equation}
Since $X_1$ does not affect the block structure, its action can be ignored for our purposes and can in fact be reabsorbed as we discussed for the matrix $\Sigma$ in \eqref{eq:gen_V}. We therefore focus on the transformation generated by $X_2$, which we expand as
\begin{equation}
\label{eq:series}
e^{iX_2} \,(C_{\alpha \beta}) \, e^{-iX_2}
=
\sum_{n=0}^{\infty} \frac{\qty[iX_2,(C_{\alpha \beta})]_n}{n!},
\end{equation}
where $[X,Y]_n = \overbrace{[X,\ldots,[X,[X,Y]]]}^{n\ \text{times}}$ denotes the $n$-fold nested commutator and $[X,Y]_0 = Y$.
Writing explicitly the term $n=1$ of the series ($n=0$ is automatically solved), we obtain
\begin{equation}
\label{eq:first_cond}
[X_2,(C_{\alpha \beta})]=
\begin{pmatrix}
    X^{(12)}(C_{\alpha \beta})^{(21)}-(C_{\alpha \beta})^{(12)}X^{(21)} & 0 \\
    0 & X^{(21)}(C_{\alpha \beta})^{(12)}-(C_{\alpha \beta})^{(21)}X^{(12)}
\end{pmatrix}.
\end{equation}
Imposing that the diagonal blocks vanish forces the entire commutator to vanish. In this case, all higher-order nested commutators in \eqref{eq:series} also vanish, ensuring that the full series has vanishing block-diagonal components and thus satisfies the constraint.
Therefore, it suffices to solve the following set of linear equations:
\begin{align}
X^{(12)}(C_{\alpha \beta})^{(21)}-(C_{\alpha \beta})^{(12)}X^{(21)}&=0, \label{eq:cons_1}\\
X^{(21)}(C_{\alpha \beta})^{(12)}-(C_{\alpha \beta})^{(21)}X^{(12)}&=0,\label{eq:cons_2}
\end{align}
for every $\alpha$ and $\beta$, which run over the elements of the operator basis used to decompose the commutators of the initial strategy  (and in principle the choice of basis can be different for the two sets of equations). 
These equations impose constraints on the coefficients of the matrix $X_2$. 
Once a solution for $X_2$ is obtained, we can obtain $U$ through exponentiation. 
Since the matrix $X_2$ is block anti-diagonal, we have a closed expression for its exponential, in terms of cosines and sines.

\paragraph{Allowed transformations} By calling $x_{ij}$ the variables of $X^{(12)}$ and using \eqref{eq:comm_comp},
the constraints \eqref{eq:cons_1} and \eqref{eq:cons_2} become
\begin{align}
    &(x_{i\alpha}^*-x_{i \beta}^*)c_{j \alpha}^*c_{j \beta}+(x_{j \alpha}-x_{j \beta})c_{i \alpha}^*c_{i \beta}=0 \ , \\
    &(x_{\alpha j}-x_{ \beta j})c_{\alpha i}^*c_{ \beta j}+(x_{\alpha i}^*-x_{ \beta i}^*)c_{\alpha j}^*c_{ \beta j}=0 \ ,
\end{align}
where $c^{(12)}_{jk}\equiv c_{jk}$.
The first immediate results are obtained by choosing $i=j$, which gives
\begin{equation}
\label{eq:constraint_real}
    \Re(x_{i \alpha})=\Re(x_{i \beta})\,, \ \qquad \Re(x_{\alpha i})=\Re(x_{\beta i})\,,
\end{equation}
namely the real part of $x_{ij}$ is constant in the entire matrix, and it is a free parameter.
Hence, the previous equations can be written in terms of the imaginary parts, let us use the letter $y$ to denote them:
\begin{align}
    &(y_{i\alpha}-y_{i \beta})c_{j \alpha}^*c_{j \beta}-(y_{j \alpha}-y_{j \beta})c_{i \alpha}^*c_{i \beta}=0 \ , \\
    &(y_{\alpha j}-y_{ \beta j})c_{\alpha i}^*c_{ \beta j}-(y_{\alpha i}-y_{ \beta i})c_{\alpha j}^*c_{ \beta j}=0 \ .
\end{align}
Focus on the first equation, and consider a $j$ for which the coefficients $c_{j\alpha}^*c_{j\beta}$ are non zero. We find
\begin{equation}
    y_{i\alpha}-y_{i \beta}=\frac{y_{j \alpha}-y_{j \beta}}{c_{j\alpha}^*c_{j\beta}}c_{i \alpha}^*c_{i \beta} \ .
\end{equation}
In order to satisfy this equation, the fraction must be independent of $j$, and equal to a parameter function only of $\alpha$ and $\beta$, call it $\lambda_{\alpha \beta}$. 
Repeating the same procedure also for the second equation, we finally get:
\begin{equation}
\label{eq:constraint_imaginary_text}
    y_{i \alpha}-y_{i \beta}=\lambda_{\alpha\beta}c_{i \alpha}^*{c_{i \beta}} \ , \qquad
    y_{\alpha i}-y_{\beta i}=\lambda'_{\alpha\beta}c_{\alpha i}^*c_{\beta i}
\end{equation}
for some parameters $\lambda_{\alpha \beta}$ and $\lambda'_{\alpha \beta}$, which is analogous to the requirement we found for the real part.

\begin{figure}
    \centering
    \includegraphics[width=0.7\linewidth]{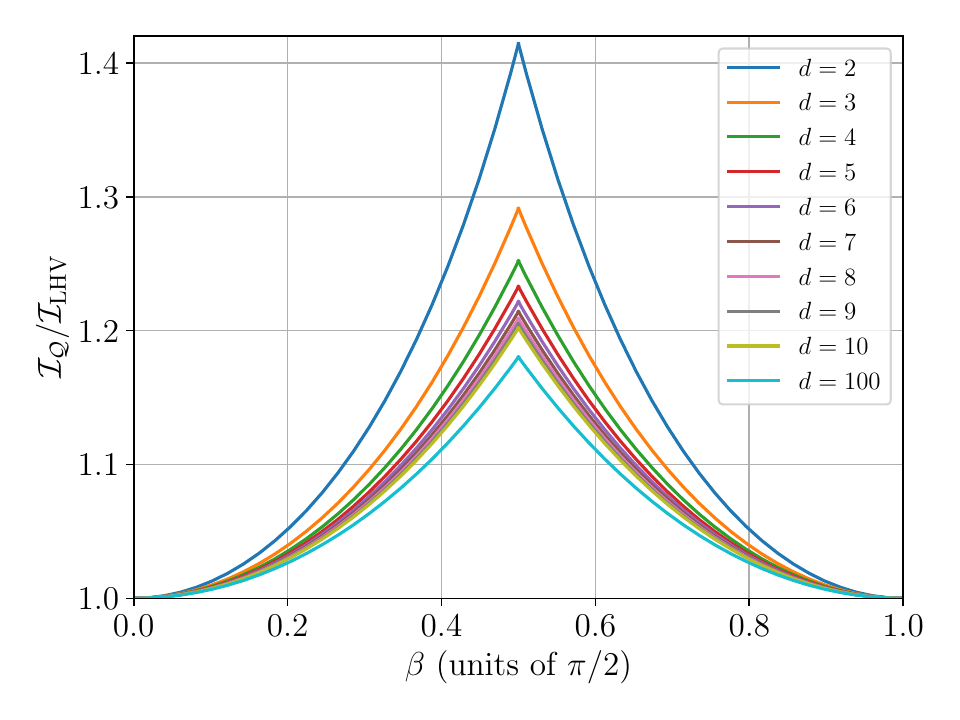}
    \caption{Ratio between quantum bound and local bound of the generalized SATWAP operator \eqref{eq:gen_satwap_exp} for different values of $\beta$ and $d$.}
    \label{fig:betaratios}
\end{figure}

\paragraph{Bell operators} Conditions \eqref{eq:constraint_imaginary_text} can be explicitly solved (see Appendix \ref{app:satwap}). 
It is simpler to represent the result by performing a permutation of the basis, and moving from the one in which the initial vector is made of $(B_0^{(1)}, \dots, B_{d-1}^{(1)},B_0^{(2)}, \dots, B_{d-1}^{(2)})$ to $(B_0^{(1)}, B_{0}^{(2)},\dots, B_{d-1}^{(1)}, B_{d-1}^{(2)})$. 
In this basis, the unitary matrix is a direct sum:
\begin{equation}
    U=\bigoplus_{k=1}^{d-1}
    \begin{pmatrix} 
    \cos \beta  & i  e^{-i \pi k / d} \sin\beta \\ ie^{i \pi k / d} \sin \beta & \cos\beta \end{pmatrix}
\end{equation}
where we assume the interval $0\le \beta \le \pi/2$. 
When $\beta=0$ or $\beta=\pi/2$ the operators collapse
to a combination of operators of the first or second input, respectively.
The transformed Bell operator is
\begin{equation}
\label{eq:gen_SATWAP_op}
\begin{split}
    S=\sum_{k=1}^{d-1}&\qty[{A}_k^{(1)}-e^{i\beta \left( 1 - \frac{2k}{d} \right)} \qty(\cos \beta  {B}_k^{(1)}-i e^{-i\pi k/d}\sin \beta {B}_k^{(2)})]^2 +\\
    &+\qty[{A}_k^{(2)}-e^{i\beta \left( 1 - \frac{2k}{d} \right)} \qty(-i e^{i\pi k/d}\sin \beta {B}_k^{(1)}+\cos \beta {B}_k^{(2)})]^2 \ , 
\end{split}
\end{equation}
where ${A}_k^{{(x)}}$ and ${B}_k^{{(y)}}$ are unitary operators with eigenvalues $\omega^{jk}$, with $j=0,\dots, d-1$.
The bound $\expval{S}=0$ is saturated by choosing $\iOP{A}_j^{(x)}$ as in \eqref{eq:Alice_CGLMP} and
\begin{equation}
    \begin{aligned}
    {\iOP{B}_k}^{(1)} &= \sum_{j=0}^{d-1} \omega^{jk} \ketbra{u_j(\beta)}{u_j(\beta)} \\
    \ket{u_j(\beta)} &= \frac{1}{\sqrt{d}} \sum_{q=0}^{d-1} \omega^{-\left( j - \frac{1}{4} - \frac{\beta}{\pi} \right) q} \ket{q} 
    \end{aligned} \ ,  \qquad
    \begin{aligned}
     {\iOP{B}_k}^{(2)} & = \sum_{j=0}^{d-1} \omega^{jk} \ketbra{v_j(\beta)}{v_j(\beta)} \\
     \ket{v_j(\beta)} &= \frac{1}{\sqrt{d}} \sum_{q=0}^{d-1} \omega^{-\left( j - \frac{3}{4} - \frac{\beta}{\pi} \right) q } \ket{q}  
    \end{aligned}
    \ . 
\end{equation}
By expanding the SOS and multiplying by $1/2$ we obtain
\begin{equation}
\label{eq:gen_satwap_exp}
    \begin{split}
        I\equiv\sum_{k=1}^{d-1} e^{-i\beta\left(1 - \frac{2k}{d}\right)} \Big[& \cos\beta {A}_k^{(1)} {B}_{d-k}^{(1)} + i e^{i\pi k/d} \sin\beta {A}_k^{(1)} {B}_{d-k}^{(2)} \\
        &+ i e^{-i\pi k/d} \sin\beta {A}_k^{(2)} {B}_{d-k}^{(1)} + \cos\beta {A}_k^{(2)} {B}_{d-k}^{(2)} \Big]\,,
    \end{split}
\end{equation}
where we used $(B^{(y)}_k)^\dagger=B^{(y)}_{d-k}$. 
The quantum bound of $I$ is $\mathcal{I}_Q=2(d-1)\openone$, while the local bound is
\begin{equation}
    \mathcal{I}_{\rm LHV} = 
\begin{cases} 
\frac{\sin(2\beta)}{2} \qty[ 2\cot\left(\frac{\beta}{d}\right) - \cot\left(\frac{\beta + \pi/2}{d}\right) + \cot\left(\frac{\pi/2 - \beta}{d}\right) ] - 2 & \text{for } 0 \le \beta \le \pi/4 \\
\\
\frac{\sin(2\beta)}{2} \qty[ \cot\left(\frac{\beta}{d}\right) + 2\cot\left(\frac{\pi/2 - \beta}{d}\right) - \cot\left(\frac{\pi - \beta}{d}\right) ] - 2 & \text{for } \pi/4 < \beta \le \pi/2 
\end{cases} \ .
\end{equation}
The family of inequalities \eqref{eq:gen_satwap_exp} generalizes the SATWAP inequality, recovered in the special case $\beta=\pi/4$. 
In Fig. \ref{fig:betaratios} we show the ratio between the quantum and local bounds of the Bell operator \eqref{eq:gen_satwap_exp} as a function of $\beta$ and for several values of $d$.

\section{Conclusions and outlook}
\label{sec:conclusions}
In this work, we addressed the problem of determining Tsirelson and local bounds for arbitrarily high-dimensional systems, in the case of two separated parties. 
We presented a systematic derivation of 
Bell operators whose Tsirelson bounds are saturated by bipartite symmetric strategies and maximally entangled states in arbitrary dimensions.
Specifically, we represent Bell parameters $S$ in terms of quadratic forms and, by applying suitable transformations, derive Tsirelson and local bounds for which an optimal strategy is guaranteed to exist. 
This existence is ensured by the fact that the transformations map sets of normal commuting operators to other sets of normal commuting operators. 
Furthermore, these transformations ensure the SOS remains observable within a Bell scenario, implying that the resulting quantum bounds are suitable for device-independent protocols, such as DI quantum key distribution or DI quantum random number generation.
As examples, in our discussions we derived several known results, both for qubit and qudit strategies, and extended them with new families of inequalities.

This work provides a foundation upon which several promising research directions and open questions can be explored.
First, an important open question concerns the completeness of this method in characterizing the quantum boundary. Although the recovery of several known and novel results suggests considerable generality, establishing the formal limits of our transformation-based approach remains an important challenge.
Moreover, we have not yet investigated sufficient or necessary conditions on Alice’s measurements and unitary transformations $U$ under which the Tsirelson bounds differ from the local bounds.

Beyond these considerations, future work may proceed along the following directions:
\begin{itemize}
    \item Deriving SOS saturated by optimal strategies using non maximally entangled states.
    \item Extending the framework to systems involving more than two parties (like, for example, those considered in \cite{Baccari:2018wjo,Santos:2022oax,Makuta:2020gjh}).
    \item Systematically investigating which bounds possess robust self-testing properties.
    \item Identifying specific device-independent quantum protocols using our results.
\end{itemize}
We are planning to discuss  some aspects of multipartite scenarios and self-testing properties in following works. 
Here we  briefly discuss some ideas and problems that arise for non-maximally entangled states.

\paragraph{Non-maximally entangled states}
Our current construction begins with a trivial Bell operator, constructed such that a maximally entangled state $\ket*{\phi^+}$ naturally resides in its kernel. 
This construction relied on the fundamental relation for maximally entangled states:
\begin{equation}
\label{eq:trick_max_ent}
    (A \otimes \openone)\ket{\phi^+} = (\openone \otimes A^T)\ket{\phi^+}
\end{equation}
which allows us to saturate a bound of $0$ using normal operators on both Alice's and Bob's sides.
For a full-rank non-maximally entangled state $\ket{\psi} = \sum_i \lambda_i \ket{ii}$ an analogous relation exists.
Define the diagonal matrix of Schmidt coefficients $\Lambda = \sum_i \lambda_i \ketbra{i}{i}$. 
Then for every Alice operator $A$, there exists a Bob operator $B$ (again the transpose of Alice's) such that
\begin{equation}
    (A \otimes \openone - \openone \otimes \Lambda A^T \Lambda^{-1})\ket{\psi} = 0 \ .
\end{equation}
While this provides a mathematical identity to "move" an operator from Alice to Bob, it introduces some problems. 
The effective operator $B_{ef}=\Lambda A^T \Lambda^{-1}$ appearing on Bob's side is generally not normal for non-maximally entangled states. 
Consequently, we cannot simply construct a physical Bell operator in the form $S = (A - B_{ef})^\dagger (A - B_{ef})$ where both $A$ and $B_{ef}$ represent standard local measurements. 
The non-normality of $B_{ef}$ implies that it does not correspond to a single physical observable in the Bell scenario.
For this reason, all the discussion presented in this paper cannot be directly applied. 
Moreover, it could be the case that the sum-of-squares method is not the optimal choice for deriving Tsirelson bounds for non-maximally entangled states  \cite{Gigena2025}.
Still, the central question and idea remain: can we construct general, physically valid Bell operators and Tsirelson bounds for non-maximally entangled states by applying unitary or more general transformations to an initial trivial Bell operator?

\section*{Acknowledgments}
LC thanks Flavio Baccari and Tommaso Grigoletto for useful discussions.
This work was supported by European Union's Horizon Europe research and innovation program under the project Quantum Secure Networks Partnership (QSNP), grant agreement No 101114043. 
Views and opinions expressed are however those of the authors only and do not necessarily reflect those of the European Union or European Commission-EU. 
Neither the European Union nor the granting authority can be held responsible for them.

\printbibliography[heading=bibintoc]

\appendix

\section{Linearly dependent qubit strategies}
\label{sec:ind_qubit}
In this appendix, we investigate the effect of linear dependencies among Alice's operators on the set of transformations that yield non-trivial Bell operators. 
We concentrate on the qubit case, for which the constraints on the unitary transformations $U$ were derived in the main text:
\begin{equation}
    \sum_{k=1}^m \qty(e^{-i \theta_j}U_{jk} - e^{i \theta_j}U_{jk}^*) \iOP{B}_k = 0.
\end{equation}
If Alice's operators are linearly independent (e.g., the three Pauli matrices) the same holds for $\iOP{B}_k$, and the only way the sum can equal zero is if every individual coefficient is exactly zero. This means that $e^{-i \theta_j}U_{jk}$ must be the components of a real matrix. 
In general, however, the initial set of $m$ matrices may have $n$ independent linear dependencies, so that there are $n$ different vectors $v^{(1)}, v^{(2)}, \dots, v^{(n)}$ such that, for any of them,
\begin{equation}
\label{eq:lin_dep}
    \sum_{k=1}^m v_k^{(\ell)}  \iOP{B}_k  = 0 \ , \quad \text{for } \ell \in \{1, \dots, n\} \ .
\end{equation}
This allows for a non-trivial imaginary part:
\begin{equation}
    U_{jk} = e^{i\theta_j}\left(R_{jk} + i \,\text{Im}\qty[\sum_{\ell=1}^n \eta_{j\ell} v_k^{(\ell)}]\right) \ , 
\end{equation}
where $R_{jk}$ is the purely real part, and $\eta_{j \ell}$ are completely arbitrary parameters.
This last degree of freedom acts like a sort of "gauge" term, in the sense that it does not affect the final expression of Bob's optimal strategy, thanks to \eqref{eq:lin_dep}.
However, they may enter in the actual expression of the Bell operator .

When requiring $U$ to be unitary, if $\eta_{j\ell}=0$ (which is always a valid solution) then (ignoring the phases) $U$ must be orthogonal, as discussed in the main text. 
If, instead, there are linear dependencies and it's possible to choose
$\eta_{j\ell} \neq 0$, the requirement of unitarity is more involved. 
In this case, the matrix $U$ can  be written compactly as:
    \begin{equation}
    U = R + iN    
    \end{equation}
    up to a irrelevant phase-diagonal matrix
    and where $N_{jk} = \text{Im}\qty[\sum_{\ell=1}^n \eta_{j\ell} v_k^{(\ell)}]$.
    For $U$ to be a unitary matrix, it must satisfy $U^\dagger U = \openone$:
    \begin{equation}
        U^\dagger U = (R^T - iN^T)(R + iN)= \openone \ . 
    \end{equation}
    Since both $R$ and $N$ are strictly real matrices, we find the conditions
    \begin{equation}
    \label{eq:qubit_lin_indip}
        R^T R + N^T N = \openone \ , \qquad R^T N - N^T R = 0 \ . 
    \end{equation}
    So, when there is linear dependence between the initial operators, the general solution also admits a non-orthogonal $R$, with a non-trivial relationship between the real part $R$ and the arbitrary parameters in $\eta$ and $v$.

\section{Restrictions for MUBs}
\label{sec:MUBs}
The number of inputs on Bob's side in \eqref{eq:S_rot_d} could, in principle, be reduced using the commutation properties of the operators in \eqref{eq:transf_op_qudit}, but this cannot be freely done for a generic pair of MUBs, as we now show. 
Suppose Alice's initial strategy consists of the measurement projectors associated with two mutually unbiased bases. 
We will restrict our analysis to a specific standard choice. 
To do so, we introduce the generalized Pauli operators $X$ and $Z$ acting on a local $d$-dimensional Hilbert space, defined by their action on the computational basis $\{|0\rangle, |1\rangle, \dots, |d-1\rangle\}$:
\begin{equation}
    X\ket*{j} = \ket*{(j+1) \pmod d}, \qquad Z\ket*{j} = \omega^j\ket*{j}
\end{equation}
where $\omega = e^{2\pi i / d}$. 
These operators satisfy the Weyl commutation relation:
\begin{equation}
    Z^k X^m = \omega^{km} X^m Z^k \ . 
\end{equation}
For the MUBs in Alice's strategy, we choose the eigenbases of $Z$ and $X$, and restrict our focus to the case where the dimension $d$ is a prime number.

Suppose we want to construct two operators, in our notation they would be $\fOP{\Pi}_1$ and $\fOP{\Pi}_2$, which are linear combinations of the projectors of these two MUBs. We can take them to be traceless.
We are saying that they can be written as
\begin{equation}
    \fOP{\Pi}_1=Z_1+X_1 \qquad \fOP{\Pi}_2=Z_2+X_2
\end{equation}
where $Z_j$ are some combinations of the projectors of the first basis, and $X_j$ of the second basis. 
Because the $d^2$ matrices $X^m Z^k$ form a complete orthogonal basis for the space of operators, any traceless operator diagonal in the first basis can be written purely as powers of $Z$, and any traceless operator diagonal in the second basis can be written purely as powers of $X$:
\begin{equation}
        Z_1 = \sum_{k=1}^{d-1} a_k Z^k \ , \qquad 
        X_1 = \sum_{m=1}^{d-1} b_m X^m \ , \qquad
        Z_2 = \sum_{k=1}^{d-1} c_k Z^k \ , \qquad
        X_2 = \sum_{m=1}^{d-1} d_m X^m \ .
\end{equation}

We require the mixed operators to commute: $[Z_1 + X_1, Z_2 + X_2] = 0$, which simplifies to $[Z_1, X_2] = [Z_2, X_1]$.
Let's compute the commutator $[Z_1, X_2]$:
\begin{equation}
    [Z_1, X_2] = \sum_{k,m} a_k d_m [Z^k, X^m] = \sum_{k,m} a_k d_m (\omega^{km} - 1) X^m Z^k  \ . 
\end{equation}
Evaluating $[Z_2, X_1]$ in the same way and equating the two expressions, we obtain:
\begin{equation}
    \sum_{k,m} a_k d_m (\omega^{km} - 1) X^m Z^k = \sum_{k,m} c_k b_m (\omega^{km} - 1) X^m Z^k \ .
\end{equation}
Because the operators $X^m Z^k$ are linearly independent basis vectors, their coefficients must match exactly term-by-term. Therefore, for every $k$ and $m$:
\begin{equation}
    a_k d_m (\omega^{km} - 1) = c_k b_m (\omega^{km} - 1) \ .
\end{equation}
Since $d$ is prime, and $k, m$ are strictly between $1$ and $d-1$, the product $km$ is never a multiple of $d$. 
Therefore, $\omega^{km}$ is never equal to $1$.
Because $(\omega^{km} - 1) \neq 0$, we can safely divide both sides by it, and we are left with:
\begin{equation}
    a_k d_m = c_k b_m \ .
\end{equation}
If one of the two final operators is a combination of only one set of projectors, say $d_m=0$, then this implies that the other operator must also be a combination of the same set ($b_m=0$).
Suppose instead $d_m \neq 0$. Then we divide by it:
\begin{equation}
   a_k = \left(\frac{b_m}{d_m}\right) c_k \ .
\end{equation}
Let $\lambda = b_m / d_m$. We now have $a_k = \lambda c_k$ for all $k$, which dictates $Z_1 = \lambda Z_2$. 
Similarly, $X_1 = \lambda X_2$.
Therefore, in $d$ dimensions with $d$ prime, linear combinations of the projectors of the MUBs associated with $Z$ and $X$ commute if and only if they are proportional.

These findings restrict the scenarios that can be considered when looking for sum-of-squares decompositions where one party's strategy consists of a pair of generic mutually unbiased bases in arbitrary dimensions \cite{Tavakoli:2020zlz}.
Globally, the two sets of projectors of two generic MUBs span a $2d-1$ dimensional space. 
Applying the unitary transformation yields $2d$ normal operators, which define $2d$ distinct inputs on Bob's side. 
To reduce the number of these inputs and simplify the scenario, some of these operators must commute. 
For this to hold for a \textit{generic} pair of MUBs in an \textit{arbitrary} dimension, the commuting operators must be proportional.
However, no more than two operators can be proportional, as exceeding this limit would reduce the dimension of the generated space too much.

\section{A $d$-dimensional local bound}
\label{app:d_local_bound}
In this appendix we compute the local bound of the operator
\begin{equation}
    2\openone - S=\sum_{k=1}^{d} \Bigg[-B_k^2 - B_{k+d}^2 + \sqrt{2}\left( \Lambda_k^{(1)} +  \Lambda_k^{(2)}\right)B_k +\sqrt{2}\left(\Lambda_k^{(1)} - \Lambda_k^{(2)}\right)B_{k+d} \Bigg] 
\end{equation}
with $d >2$, introduced in Sec.\ \ref{sec:qudit_orth}. Here, $\Lambda_k^{(x)}$ are two sets of projectors and Bob's operators have $d-2$ zero eigenvalues while the others are:
\begin{equation}
\label{eq:lambda_bob}
\lambda_{\pm}=
\begin{cases}
\lambda_{\pm}^{(1)}\equiv\frac{1}{\sqrt{2}} \left(1\pm \frac{1}{\sqrt{d}} \right) \qquad &1 \le k \le d \\
\lambda_{\pm}^{(2)}\equiv \pm \frac{1}{\sqrt{2}} \sqrt{1- \frac{1}{d}} \qquad &d+1 \le k \le 2d 
\end{cases}
\end{equation}

In the  local scenario,  we treat all operators as commuting scalar variables.
For deterministic classical strategies, Alice's measurement outputs exactly one deterministic value for each input. 
This means replacing the operator $\Lambda_k^{(x)}$ with a boolean variable $ \lambda_k^{(x)}\in \{0, 1\}$, and for a fixed $x$, exactly one $k$ evaluates to $1$ while the rest are $0$. 
Bob's outcomes instead are encoded in the variable $b_k$ and $b_{k+d}$, restricted to the eigenvalues \eqref{eq:lambda_bob} and the $d-2$ null eigenvalues.
We want to maximize the scalar sum:
\begin{equation}
    I_{\rm LHV} = \sum_{k=1}^d \left[ -b_k^2 + \sqrt{2}\qty(\lambda_k^{(1)}+\lambda_k^{(2)}) b_k - b_{k+d}^2 + \sqrt{2}\qty(\lambda_k^{(1)}-\lambda_k^{(2)}) b_{k+d} \right] \ .
\end{equation}
Because $\sum_k \lambda_k^{(x)} = 1$, Alice has two distinct deterministic behaviors to consider over the inputs:
\begin{itemize}
    \item Case A: Alice outputs the same $k$ for both input $1$ and input $2$.
    For that specific $k$, $\lambda_k^{(1)} = \lambda_k^{(2)} = 1$.
    To maximize Bob's side, we assign $b_{k+d} = 0$ and pick the largest available eigenvalue for $b_k$, which is $\lambda_+ = \frac{1}{\sqrt{2}}(1 + \frac{1}{\sqrt{d}})$.
    Substituting these in gives the maximum for this scenario:
    \begin{equation}
        \mathcal{I}_{\rm LHV}= -\lambda_+^2 + 2\sqrt{2}\lambda_+ = \frac{3}{2} + \frac{1}{\sqrt{d}} - \frac{1}{2d} \ .
    \end{equation}
    \item Case B: Alice outputs $k_1$ for input $1$ and $k_2$ for input $2$ ($k_1 \neq k_2$).
    For the term $k_1$ in the sum, we pick $b_{k_1} = \lambda_+^{(1)}$ and $b_{k_1+d} = +\frac{1}{\sqrt{2}}\sqrt{1 - 1/d}$. 
    The maximum for this term is $\sqrt{1 - 1/d}$. For $k_2$, the maximum similarly evaluates to $+\sqrt{1 - 1/d}$.
    Summing both contributions yields the maximum for this case:
    \begin{equation}
        \mathcal{I}_{\rm LHV} = 2\sqrt{1 - \frac{1}{d}} \ .
    \end{equation}
\end{itemize}
The local bound is  whichever strategy yields the largest number:
\begin{equation}
    \mathcal{I}_{\rm LHV} = \max \left[ \frac{3}{2} + \frac{1}{\sqrt{d}} - \frac{1}{2d}, 2\sqrt{1 - \frac{1}{d}} \right] \ . 
\end{equation}
Which one is larger depends  on the dimension $d$: For lower dimensions ($d \le 5$), the first strategy  is the maximum. For higher dimensions ($d \ge 6$), the second one.

\section{Generalization SATWAP}
\label{app:satwap}
In this appendix, as a case study, we want to show the details of the derivation and generalization of SATWAP results \cite{SATWAP2017}. 
Here, we present all the details not explicitly written in Sec.\ref{sec:d_m2}.

From the discussion in the main text, we need to solve, in $y$, the equations:
\begin{equation}
\label{eq:constraint_imaginary}
    y_{i \alpha}-y_{i \beta}=\lambda_{\alpha\beta}c_{i \alpha}^*{c_{i \beta}} \ , \qquad
    y_{\alpha i}-y_{\beta i}=\lambda'_{\alpha\beta}c_{\alpha i}^*c_{\beta i}
\end{equation}
for some parameters $\lambda_{\alpha \beta}$ and $\lambda'_{\alpha \beta}$ and where
\begin{equation}
    c_{jk}=\frac{2}{d}\left(\frac{1}{1-\omega^{k-j-1/2}}\right) \ .
\end{equation}
The basis transition matrix satisfies some properties which significantly simplify the problem, namely
\begin{align}
\label{eq:overlap_cond}
     c_{i \alpha}^*{c_{i \beta}}=\xi_{\alpha \beta}\left(h_{\alpha i}-h_{\beta i} \right) \\
     c_{\alpha i}^*{c_{\beta i}}=\xi'_{\alpha \beta}\left(h'_{\alpha i}-h'_{\beta i} \right) 
\end{align}
with
\begin{equation}
\begin{aligned}
     \xi_{\alpha \beta} &= \frac{4}{d^2} \frac{1}{1 - \omega^{\beta-\alpha}} \\
   h_{\alpha j} &= \frac{1}{1 - \omega^{j-\alpha+1/2}}
\end{aligned}    
    \ , \qquad \qquad 
\begin{aligned}    
 \xi'_{\alpha \beta} &= \frac{4}{d^2} \frac{1}{1 - \omega^{\alpha-\beta}} \\
h'_{\alpha j}& = \frac{1}{1 - \omega^{\alpha-j+1/2}}=h_{j\alpha}
\end{aligned} \ .
\end{equation}
Let us now focus on the left equation in \eqref{eq:constraint_imaginary}. Using the factorization property the (trivial) identity
\begin{equation}
(y_{i \alpha}-y_{i \beta })+(y_{i \beta}-y_{i \gamma })+(y_{i \gamma}-y_{i \alpha})=0 
\end{equation}
can be written as
\begin{equation}
\label{eq:lin_ind}
    h_{\alpha i}(\lambda_{\alpha \beta} \xi_{\alpha \beta} - \lambda_{\gamma \alpha} \xi_{\gamma \alpha})+h_{\beta i}(\lambda_{\beta \gamma} \xi_{\beta \gamma}-\lambda_{\alpha \beta} \xi_{\alpha \beta})  + h_{\gamma i}(\lambda_{\gamma \alpha} \xi_{\gamma \alpha}-\lambda_{\beta \gamma} \xi_{\beta \gamma})= 0 \ .
\end{equation}
Since the $h_{\alpha i}$ form a linearly independent basis, we obtain that the coefficients in equation \eqref{eq:lin_ind} must vanish, namely that $\lambda_{\alpha \beta}\xi_{\alpha \beta}=C$, for a constant $C$. Substituting this back,
\begin{equation}
    y_{i j}=y_{i 0}+\frac{C}{\xi_{j0}}c_{i j}^*{c_{i 0}}=y_{i0}+C(h_{ji}-h_{0i}) \ . 
\end{equation}
The first column, $y_{i0}$, can be arbitrarily chosen ($d$ variables), together with the (real) scaling constant $C$. The second constraint in \eqref{eq:constraint_imaginary} can be tackled in exactly the same way, leading to
\begin{equation}
    y_{j i}=y_{0 i}+\frac{C'}{\xi'_{j0}}c_{ j i}^*{c_{0 i}}=y_{0i}+C'(h'_{ji}-h'_{0i})=y_{0i}+C'(h_{ij}-h_{i0})
\end{equation}
with a constant $C'$, and this time we can arbitrarily choose the first row, $y_{0i}$. 
However, we now need to simultaneously solve these two sets, so that
$C$ and $C'$ are not independently free parameters. 
To find a closed form for any $y_{ij}$, we first express the boundary elements $y_{i0}$ and $y_{0j}$ in terms of the single scalar $y_{00}$. By choosing $i=0$ in both equations, we have:
\begin{align}
    y_{0j} &= y_{00} + C(h_{j0}-h_{00}) \ , \\
    y_{j0} &= y_{00} + C'(h_{0j}-h_{00}) \ .
\end{align}
We can now plug back, obtaining two closed expressions
\begin{align}
    y_{ij} &= y_{00}+C'(h_{0i}-h_{00})+C(h_{ji}-h_{0i}) \ , \\
    y_{ij} &= y_{00}+C(h_{j0}-h_{00})+C'(h_{ji}-h_{j0}) \ ,
\end{align}
which imply the consistency condition:
\begin{equation}
    C' \left( h_{0i}-h_{00}-h_{ji}+h_{j0}\right) = C \left( h_{j0}-h_{00}-h_{ji}+h_{0i} \right) \qquad \Rightarrow \qquad C=C' \ .
\end{equation}
Therefore, the general elements of the matrix are
\begin{equation}
\label{eq:imag_const}
     y_{ij} = y_{00}+C(h_{ji}-h_{00}) \ .
\end{equation}
Note that the quantity $(h_{ji}-h_{00})$ is pure imaginary, so also $C$ must be imaginary to satisfy \eqref{eq:imag_const}.

The $y$'s we just derived are the imaginary parts of the elements of the matrix $X^{(12)}$, introduced in the main text.
We already argued, around Eq.\ \eqref{eq:constraint_real}, that the real parts of these elements are constants. 
Combining them to form the full complex matrix elements, and rescaling $i C \to b$, with $b \in \mathbb{R}$, we find
\begin{equation}
     x_{jk} = a+b h_{kj}
\end{equation}
where $a=x_{00}+iy_{00}-iCh_{00}$ is a generic complex constant and $h_{kj}=1/(1-\omega^{j-k+1/2})$. 
Note that the constant term corresponds to a shift in the direction of the identity.

\subsubsection*{Change of basis}
As we said, the variables $x_{jk}$ are elements of a generic matrix $X^{(12)}$, and we need to consider the exponential of
\begin{equation}
\label{eq:X2_app}
    X_2=
    \begin{pmatrix}
    0 & X^{(12)} \\
    {X^{(12)}}^\dagger & 0
    \end{pmatrix} \ .
\end{equation}
Being block off-diagonal, the exponential can be computed. 
However we can make the problem even easier by changing basis and moving to the basis of the normal operators used to saturate the SATWAP, by computing $
    M X_2 M^{-1}$  with $M=\bigoplus_xM^{(x)}$ and 
\begin{equation}
    M^{(x)}_{jk} = \omega^{jk} \ . 
\end{equation}
The transformation of each block is given by $P = M^{(x)} X^{(12)} (M^{(x)})^{-1}$. Using that $(M^{(x)})^{-1}_{kn} = \frac{1}{d} \omega^{-kn}$, the elements of the matrix $P$ are:
\begin{equation}
    p_{mn} = \sum_{j=0}^{d-1} \sum_{k=0}^{d-1} M^{(x)}_{mj} x_{jk} (M^{(x)})^{-1}_{kn} = \frac{1}{d} \sum_{j=0}^{d-1} \sum_{k=0}^{d-1} \omega^{mj} \left( a + \frac{b}{1 - \omega^{j-k+1/2}} \right) \omega^{-kn} \ . 
\end{equation}
We can separate this into a constant term $p^{(a)}_{mn}$ and a shifted term $p^{(b)}_{mn}$, corresponding to the two elements in the brackets. 
For the constant term $a$, we use the orthogonality of the roots of unity, $\sum_{k=0}^{d-1} \omega^{ck} = d \delta_{c,0}$:
\begin{equation}
    p^{(a)}_{mn}= \frac{a}{d} \left( \sum_{j=0}^{d-1} \omega^{mj} \right) \left( \sum_{k=0}^{d-1} \omega^{-kn} \right)= a d \delta_{m,0} \delta_{n,0} \ .
\end{equation}
For the term involving $b$, we substitute $s = j-k$ (so $j = k+s$). Taking the indices modulo $d$:
\begin{align}
    p^{(b)}_{mn} &= \frac{b}{d} \sum_{k=0}^{d-1} \sum_{s=0}^{d-1} \omega^{m(k+s)} \frac{1}{1 - \omega^{s+1/2}} \omega^{-kn}= \frac{b}{d} \left( \sum_{k=0}^{d-1} \omega^{k(m-n)} \right) \left( \sum_{s=0}^{d-1} \frac{\omega^{ms}}{1 - \omega^{s+1/2}} \right) \\
    &= b \delta_{m,n} \sum_{s=0}^{d-1} \frac{\omega^{ms}}{1 - \omega^{s+1/2}} \ . 
\end{align}
The factor $\delta_{m,n}$ proves that the resulting matrix is diagonal. Let $x = \omega^{-1/2} = e^{-i\pi/d}$. The diagonal sum becomes:
\begin{equation}
    T_m = x \sum_{s=0}^{d-1} \frac{(\omega^s)^m}{x - \omega^s} \ .
\end{equation}
Using the root of unity identity $\sum_{k=0}^{d-1} \frac{z_k^m}{x - z_k} = \frac{d x^{m-1}}{x^d - 1}$ (where $z_k = \omega^k$) and noting that $x^d = (e^{-i\pi/d})^d = -1$, we evaluate $T_m$ for two cases:

For $m = 0$:
\begin{equation}
    T_0 = x \frac{d x^{d-1}}{x^d - 1} = \frac{d x^d}{x^d - 1} = \frac{d(-1)}{-1 - 1} = \frac{d}{2}
\end{equation} \ ,

For $m \neq 0$:
\begin{equation}
    T_m = x \frac{d x^{m-1}}{x^d - 1} = \frac{d x^m}{-2} = -\frac{d}{2} (e^{-i\pi/d})^m = -\frac{d}{2} \omega^{-m/2} \ . 
\end{equation}
Combining $p^{(a)}_{mn}$ and $p^{(b)}_{mn}$, the final transformed block is a diagonal matrix with elements:
\begin{equation*}
    p_{mn} = \begin{cases} 
      \alpha  & \text{for } m = n = 0 \\
      \beta \omega^{-m/2} & \text{for } m = n \neq 0 \\
      0 & \text{for } m \neq n 
   \end{cases}
\end{equation*}
where we defined $\alpha=d(a+b/2)$ (complex) and $\beta=-d b/2$ (real).
In this way, we have identified $P$, which corresponds to the block $X^{(12)}$ from \eqref{eq:X2_app} expressed in the new basis.
The bottom left matrix in \eqref{eq:X2_app} is just the conjugate transpose.

\subsubsection*{Computation of the exponential}
The exact analytical form for the matrix exponential is:
\begin{equation}
    e^{i X_2} = \begin{pmatrix} \cos(D) & i P D^{-1} \sin(D) \\ 
    iP^\dagger D^{-1} \sin(D) & \cos(D) \end{pmatrix}
\end{equation}
where $D$ has entries $d_m = |p_{mm}|$
Because everything inside these blocks is diagonal, we can write down the exact action for any specific $2 \times 2$ subspace $m$ (coupling state $m$ with state $m+d$). For a given $m$, the $2 \times 2$ submatrix is:
\begin{equation}
    \exp \left[ i\begin{pmatrix} 0 & p_{mm} \\ p_{mm}^* & 0 \end{pmatrix} \right] = \begin{pmatrix} \cos(|p_{mm}|) & \frac{p_{mm}}{|p_{mm}|} i\sin(|p_{mm}|) \\ i\frac{p_{mm}^*}{|p_{mm}|} \sin(|p_{mm}|) & \cos(|p_{mm}|) \end{pmatrix} \ .
\end{equation}
The subspace corresponding to $m=0$ is irrelevant because, for $m=0$, the operator is the identity and the corresponding term in the SOS disappears. Since it is decoupled from the others, we can ignore it.
For $m \neq 0$,  we get
\begin{equation}
    \begin{pmatrix} 
    \cos \beta  & i  e^{-i \pi m / d} \sin\beta \\ ie^{i \pi m / d} \sin \beta & \cos\beta \end{pmatrix}
\end{equation}
where we assume the interval $0\le \beta \le \pi/2$. 
 When $\beta=0$ or $\beta=\pi/2$ the operators collapse to a combination of operators of the first or second input, respectively. Including a larger interval would just translate into a redefinition of the sign of the operators.

\subsubsection*{Transformed operators and spectral decomposition}
The final optimal strategy for Bob is identified by
\begin{align}
    \fOP{B}_{j}^{(1)}&=\cos\beta \iOP{B}_j^{(1)}+i \sin  \beta \ e^{-i \pi j/d}\iOP{B}_j^{(2)} \\
     \fOP{B}_{j}^{(2)}&=\cos \beta \iOP{B}_j^{(2)}+i \sin \beta \ e^{i \pi  j/d}\iOP{B}_j^{(1)}
\end{align}
and we should now find the spectral decomposition of these operators to plug it into the transformed expression of the Bell operator.
Recall that
\begin{equation}
\iOP{B}_j^{(x)} = \sum_{k=0}^{d-1} \omega^{kj} \ketbra*{\beta_k^{(x)}}{\beta_k^{(x)}} \\
= \frac{1}{d} \sum_{p,q=0}^{d-1} \omega^{(\frac{x}{2}-\frac{1}{4})(p-q)} \ketbra{q}{p} \left( \sum_{k=0}^{d-1} \omega^{k(j-p+q)} \right) \ . 
\end{equation}
The sum over $k$ evaluates to $d$ if $p = q + j \pmod{d}$, and $0$ otherwise. Recalling that $0 \le j < d$, we have two cases:
\begin{itemize}
    \item If $q + j < d$, then $p = q + j \implies p - q = j$. The phase is $\omega^{j/4}$.
    \item  If $q + j \ge d$, then $p = q + j - d \implies p - q = j - d$. The phase is $\omega^{(j-d)/4} = -i \omega^{j/4}$.
\end{itemize}
Let us denote $q \oplus j \equiv (q + j) \pmod{d}$. Thus, $\iOP{B}_j^{(1)}$ acts as a cyclic shift with an extra phase:
\begin{equation}
    \iOP{B}_j^{(1)} \ket{q} = c_q \omega^{j/4} \ket{q \oplus j}
\end{equation}
where $c_q = 1$ if $q < d - j$, and $c_q = -i$ if $q \ge d - j$.
To write a similar expression for $B_j^{(2)}$, notice that the states $\ket*{\beta_k^{(1)}}$ and $\ket*{\beta_k^{(2)}}$ are related by a diagonal phase operator as:
\begin{equation}
    \ket{\beta_k^{(2)}} = W \ket{\beta_k^{(1)}}  \qquad \text{with} \qquad W = \sum_{q} \omega^{q/2} \ketbra{q}{q} = \sum_{q} e^{i \pi q / d} \ketbra{q}{q} \ .
\end{equation}
This implies that $\iOP{B}_j^{(2)} = W \iOP{B}_j^{(1)} W^\dagger$ and
$\iOP{B}_j^{(2)} \ket{q}$:
\begin{equation}
    \iOP{B}_j^{(2)} \ket{q} = W \iOP{B}_j^{(1)} e^{-i \pi q / d} \ket{q} = e^{i \pi (q \oplus j)/d} e^{-i \pi q / d} c_q \omega^{j/4} \ket{q \oplus j} = e^{i \pi (q \oplus j - q)/d} \iOP{B}_j^{(1)} \ket{q} \ .    
\end{equation}
If $q + j < d$, the exponent factor is $e^{i \pi j / d}$, and we get $e^{i \pi j / d} \iOP{B}_j^{(1)} \ket{q}$. If $q + j \ge d$, the exponent factor is $e^{i \pi (j - d) / d} = -e^{i \pi j / d}$, and we get $-e^{i \pi j / d} \iOP{B}_j^{(1)} \ket{q}$.
So, as before, we have a compact form:
\begin{equation}
    \iOP{B}_j^{(2)} \ket{q} = s_q e^{i \pi j / d} \iOP{B}_j^{(1)} \ket{q}
\end{equation}
where $s_q = 1$ for $q < d - j$, and $s_q = -1$ for $q \ge d - j$.
Substituting this mapping into $\fOP{B}_j^{(1)}$, we get:
\begin{equation}
    \fOP{B}_j^{(1)} \ket{q} = \left( \cos\beta \iOP{B}_j^{(1)} + i \sin\beta e^{-i \pi j / d} \iOP{B}_j^{(2)} \right) \ket{q} = (\cos\beta + i s_q \sin\beta) \iOP{B}_j^{(1)} \ket{q} = e^{i s_q \beta} \iOP{B}_j^{(1)} \ket{q} \ .
\end{equation}
Similarly, using $\iOP{B}_j^{(1)} \ket{q} = s_q e^{-i \pi j / d} \iOP{B}_j^{(2)} \ket{q}$, we find:
\begin{equation}
    \fOP{B}_j^{(2)} \ket{q} = e^{i s_q \beta} \iOP{B}_j^{(2)} \ket{q} \ .
\end{equation}
Both the final operators are simply the original operators multiplied by a diagonal phase operator.
To find the spectral decomposition of $\fOP{B}_j^{(1)}$, we set up the eigenvalue equation $\fOP{B}_j^{(1)} \ket{w_k} = \lambda_k \ket{w_k}$ and, given its relation to the initial operators, we make the ansatz of a Fourier-like eigenstate $\ket{w_k} = \frac{1}{\sqrt{d}} \sum_q e^{i \theta_q} \ket{q}$. Applying $\fOP{B}_j^{(1)}$ gives:
\begin{equation}
    \fOP{B}_j^{(1)} \ket{w_k} = \frac{1}{\sqrt{d}} \sum_{q=0}^{d-1} e^{i \theta_q} e^{i s_q \beta} c_q \omega^{j/4} \ket{q \oplus j} = \lambda_k \frac{1}{\sqrt{d}} \sum_{p=0}^{d-1} e^{i \theta_p} \ket{p} \ .
\end{equation}
To compare the two sides, we need to consider $p = q \oplus j$ and again we have two regimes. 
Separating into the two piecewise regimes ($p \ge j$ and $p < j$), we obtain two recurrence conditions for the phases:
\begin{itemize}
    \item If $q+j<d$ then $p=q+j$. Since $q\ge 0$, it implies that $p \ge j$. In this case: 
    \begin{equation}
    \label{eq:ph1}
        e^{i \theta_{p-j}} e^{i \beta} \omega^{j/4} = \lambda_k e^{i \theta_p}
    \end{equation}
    \item If $q+j \ge d$ then $p=q+j-d$. Since $q\le d-1$, it implies $p < j$. In this case: 
    \begin{equation}
    \label{eq:ph2}
        e^{i \theta_{p-j+d}} e^{-i \beta} (-i) \omega^{j/4} = \lambda_k e^{i \theta_p}
    \end{equation}
\end{itemize}
To simultaneously solve the two equations \eqref{eq:ph1} and \eqref{eq:ph2}, we assume a linear phase ansatz $\theta_q = -K q$. Dividing the two equations to isolate $K$ yields $e^{-i K d} = i e^{2i \beta}$, which gives the allowed frequencies:
\begin{equation}
    K = \frac{2\pi}{d} \left( k - \frac{1}{4} - \frac{\beta}{\pi} \right) \quad \text{for } k = 0, 1, \dots, d-1
\end{equation}
Substituting $K$ back into either recurrence relation yields the exact eigenvalues $\lambda_k$ and gives the spectral decomposition of $\fOP{B}_j^{(1)}$:
\begin{equation}
    \fOP{B}_j^{(1)} = \sum_{k=0}^{d-1} \lambda_k \ketbra{u_k(\beta)}{u_k(\beta)}
\end{equation}
where the eigenvectors $\ket{u_k(\beta)}$ and eigenvalues $\lambda_k$ are:
\begin{align}
    \ket{u_k(\beta)} &= \frac{1}{\sqrt{d}} \sum_{q=0}^{d-1} \exp\left[ -\frac{2\pi i}{d} \left( k - \frac{1}{4} - \frac{\beta}{\pi} \right) q \right] \ket{q} \\
    \lambda_k &= \omega^{jk} \exp\left[ i\beta \left( 1 - \frac{2j}{d} \right) \right] \ .
\end{align}
Because $\fOP{B}_j^{(2)} = W \fOP{B}_j^{(1)} W^\dagger$, it shares the exact same eigenvalues $\lambda_k$, and its eigenvectors are phase-shifted by $W$:
\begin{equation}
    \fOP{B}_j^{(2)} = \sum_{k=0}^{d-1} \lambda_k \ketbra{v_k(\beta)}{v_k(\beta)}
\end{equation}
where the eigenvectors $\ket{v_k(\beta)} = W \ket{u_k(\beta)}$ are given by:
\begin{align}
    \ket{v_k(\beta)} = \frac{1}{\sqrt{d}} \sum_{q=0}^{d-1} \exp\left[ -\frac{2\pi i}{d} \left( k - \frac{3}{4} - \frac{\beta}{\pi} \right) q \right] \ket{q} \ .
\end{align}
As a quick check, setting $\beta = 0$ perfectly collapses $\ket*{u_k(\beta)}$ into $\ket*{\beta_k^{(1)}}$, $\ket*{v_k(\beta)}$ into $\ket*{\beta_k^{(2)}}$, and $\lambda_k$ into $\omega^{jk}$, recovering the original operators $\iOP{B}_j^{(1)}$ and $\iOP{B}_j^{(2)}$.
The measurements for the standard SATWAP inequality are recovered by choosing $\beta=\pi/4$.

\end{document}